\documentclass{article}
\usepackage{algorithm}
\usepackage[noend]{algpseudocode}
\DeclareUnicodeCharacter{3B1}{\ensuremath{\alpha}}
\algrenewcommand{\algorithmiccomment}[1]{$\vartriangleright$ #1}
\algrenewcommand{\algorithmicreturn}{\textbf{Return: }}
\algnewcommand\algorithmicinput{\textbf{Input: }}
\algnewcommand\Input{\State \algorithmicinput}

\usepackage{amsmath, amssymb}
\usepackage[x12names, rgb, html]{xcolor}

\definecolor{red}{HTML}{f54b1a}
\definecolor{pink}{HTML}{d19eb1}
\definecolor{orange}{HTML}{d3772e}
\definecolor{yellow}{HTML}{ebe85d}
\definecolor{green}{HTML}{0f6852}
\definecolor{lightblue}{HTML}{01abe9}
\definecolor{darkblue}{HTML}{1b346c}
\definecolor{tan}{HTML}{e5c39e}
\definecolor{darktan}{HTML}{af9e73}
\definecolor{grey}{HTML}{c3ced0}
\definecolor{darkgrey}{HTML}{9dadc4}
\definecolor{black}{HTML}{110d1b}
\definecolor{white}{HTML}{f1f8f1}

\usepackage{hyperref}
\hypersetup{
  colorlinks = true,
  linkcolor = red,
  citecolor = darkblue,
  urlcolor = lightblue
}

\usepackage[margin=1in]{geometry}

\usepackage{graphicx}
\graphicspath{{figs/}}

\def\xb{\boldsymbol{x}}
\def\yb{\boldsymbol{y}}

\def\RR{\mathbb{R}} \def\NN{\mathbb{N}}

\def\<{\langle} \def\>{\rangle}

\usepackage{graphicx} 
\usepackage{booktabs}
\usepackage{caption}
\usepackage[style=numeric, maxnames=10, backend=biber]{biblatex}
\AtEveryBibitem{\clearfield{note}}

\title{Data-efficient generation of protein conformational ensembles with backbone-to-side chain transformers}
\author{
Shriram Chennakesavalu$^{1}$ and Grant M. Rotskoff$^{1,2}$\footnote{Contact: \texttt{rotskoff@stanford.edu}}
 \\
1. Department of Chemistry, Stanford University, Stanford, CA, USA 94305 \\
2. Institute for Computational and Mathematical Engineering, \\Stanford University, Stanford, CA, USA 94305 
}
\date{\today}

\begin{document}

\maketitle

\begin{abstract}
Excitement at the prospect of using data-driven generative models to sample configurational ensembles of biomolecular systems stems from the extraordinary success of these models on a diverse set of high-dimensional sampling tasks. Unlike image generation or even the closely related problem of protein structure prediction, there are not currently data sources with sufficient breadth to parameterize generative models for conformational ensembles. To enable discovery, a fundamentally different approach to building generative models is required: models should be able to propose rare, albeit physical, conformations that may not arise in even the largest data sets. Here we introduce a modular strategy to generate conformations based on ``backmapping'' from a fixed protein backbone that 1) maintains conformational diversity of the side chains and 2) couples the side chain fluctuations using global information about the protein conformation. Our model combines simple statistical models of side chain conformations based on rotamer libraries with the now ubiquitous transformer architecture to sample with atomistic accuracy. Together, these ingredients provide a strategy for rapid data acquisition and hence a crucial ingredient for scalable physical simulation with generative neural networks. 
\end{abstract}

\section{Introduction}
Generative models offer an appealing route to efficiently sample biomolecular conformational ensembles~\cite{noe_boltzmann_2019,chennakesavalu_ensuring_2023,zheng_towards_2023}.
The high-dimensional, metastable probability distributions characteristic of thermal ensembles of large biomolecules are notoriously difficult to sample~\cite{noe_machine_2019}, requiring sophisticated importance sampling techniques~\cite{lelievre_free_2010} and high-performance computing~\cite{lindorff-larsen_how_2011}.
While generative models may reduce the expense of exploring conformational space, there are not yet scalable methods for training these models.
At present, the most powerful techniques for generative modeling have shown promise primarily on tasks where massive data sets are readily available~\cite{song_score-based_2022, albergo_building_2022, lipman_flow_2022}. 
For molecular systems, data is limited by the low spatial resolution of single-molecule measurements and the aforementioned cost of molecular dynamics (MD) simulation.
In large-scale databases of three-dimensional protein structures, like the Protein Data Bank, rare but important transition states and metastable configurations are entirely absent. 
Building generative models that do not depend on task-specific data and that can be deployed on new biomolecular systems with little or no molecule-specific adaptation remains a defining challenge for this emerging field.

To construct models that do not require task-specific data, there are two dominant strategies: First, one can use data from a closely related task that is inexpensive to acquire and carry out transfer learning. 
While this approach is appealing, designing a transfer learning strategy that does not over-constrain the model and result in negative transfer can be challenging~\cite{gerace_probing_2022, hanneke_value_2019}.
Secondly, building a model that incorporates strong physical inductive biases in the form of constraints or preconditioners can improve generalization by ensuring that generated samples respect the imposed constraints~\cite{zheng_towards_2023, lai_fp-diffusion_2023}. 
Physics-based bottom-up coarse-grained models~\cite{noid_multiscale_2008,jin_bottom-up_2022,izvekov_multiscale_2005,saunders_coarse-graining_2013} offer a particularly attractive approach to building such a prior~\cite{chennakesavalu_ensuring_2023}.  
At the same time, imposing these restrictions may impact the trainability and expressiveness of the model in ways that are unproductive~\cite{kipf_semi-supervised_2017}.
Strategies for combining these approaches to build truly transferable generative models is an area of ongoing research.

In parallel to the rapid developments of state-of-the-art generative models for image generation, an improved understanding of how to use normalizing flows~\cite{tabak_density_2010, rezende_variational_2015} for variational inference tasks has begun to emerge~\cite{gabrie_efficient_2021}.
Normalizing flows are deterministic diffeomorphisms that transform samples from a tractable probability distribution, like a Gaussian, to a given target distribution. 
Because the likelihoods can be computed explicitly and efficiently for some architectures~\cite{dinh_density_2017, durkan_neural_2019}, it is possible to train these models \emph{without data}, using a maximum likelihood objective.
Unfortunately, mode collapse, a phenomenon in which all the probability mass of the generated distribution collapses on a single mode, appears to be a fundamental issue with this \emph{data-free} training approach for metastable distributions.
Our previous work~\cite{chennakesavalu_ensuring_2023} showed that combining physics-based coarse-grained models with an invertible backmapping strategy mitigates mode collapse and enables a sampling strategy that is amenable to exact reweighting. 
However, the procedure we used for backmapping coarse-grained structures, while statistically controlled in the sense that it ensures Boltzmann statistics, is not ``task-independent,'' and the framework we employed was limited by the requirement of atomistic configurational data of the target molecule.

Throughout this work, we focus on the problem of reconstructing an atomic resolution protein structure from only information about its backbone coordinates.
Understanding and analyzing the side chain conformations available to different residue types, as well as their dependence on the backbone configuration, is an old problem. 
Early work~\cite{chanhrasekaran_studies_1970} established that the number of rotameric states available to the side chains of biological amino acids was finite, which has since prompted statistical investigations into the sequence dependence of these side chain rotameric state~\cite{bower_prediction_1997, dunbrack_jr_bayesian_1997}.
Backmapping protein structures from a coarse-grained configuration, often the backbone, bears a statistical resemblance to Levinthal's famous ``paradox''~\cite{levinthal_are_1968} which is resolved simply with the recognition that the strong correlations---arising due to the potential energy---among rotamers limits the number of thermodynamically accessible states. 
So, while the many-body energetic interactions among the atoms of a protein constrain its structure, a naive attempt to sample a configuration by independently sampling rotamers along a backbone will fail.
Indeed, direct investigations of the correlations between side chains in a protein indicate that the information propagates on a scale comparable to the extent of the protein itself~\cite{dubay_calculation_2009, dubay_fluctuations_2015, dubay_long-range_2011}.

Due to the tractable structure of rotamer distributions, several groups have pursued efforts to systematically enumerate the available states~\cite{dunbrack_rotamer_2002,dicks_exploiting_2022}.
Furthermore, these libraries have subsequently been deployed to capture the sequence dependence of side chains using relaxation algorithms~\cite{miao_rasp_2011, krivov_improved_2009, jumper_rapid_2017}. 
Recently, this topic has been revisited in the context of more expressive probabilistic models built with deep neural networks~\cite{stieffenhofer_adversarial_2020, chennakesavalu_ensuring_2023, yang_chemically_2023, jones_diamondback_2023}.
However, in large part, these efforts have not sought to add physically meaningful thermal stochastic fluctuations.

In this work, we aim to build samplers that demonstrate high levels of agreement for correlated side chain configurations with Boltzmann distributed configurations from atomistic MD.
We do so by exploiting the observation that rotamer libraries can easily be modeled at single side-chain resolution with a Gaussian mixture model, as demonstrated below.
The most challenging aspect of backmapping is then reduced to learning the coupling among these simple probabilistic models. 
This problem---given a backbone sequence, generate a sequence of side chains consistent with that backbone conformation---is very naturally suited to sequence-to-sequence models capable of capturing long-range correlations, such as transformers~\cite{vaswani_attention_2017}.
We demonstrate that---even without \textit{a priori} access to all-atom data of the target protein---learning a many-body coupling among side chain conformations from rotamer libraries with a transformer yields backmapped configurations in excellent agreement with atomistic simulations. This level of generalization lends itself to rapid enumeration of conformational states that may only interconvert on timescales inaccessible to MD.

\section{Related Work}

\paragraph{Backmapping coarse-grained representations}

A number of works have explored backmapping from a coarse-grained representation of a molecular system to atomistic resolution. 
Deterministic algorithms that construct an atomistic configuration that avoids steric clashes have been proposed for both proteins~\cite{miao_rasp_2011, jumper_rapid_2017} and DNA~\cite{walther_multi-modal_2020}. Of course, because these algorithms are deterministic, they do not recover the entropy lost when coarse-graining. An early stochastic approach used Dynamical Bayesian Networks to sample side chain conformations conditioned on backbone dihedral angles \cite{harder_rotamers_2010}.
More recently, several stochastic approaches based on generative neural networks have been proposed.
For example, Stieffenhofer et al.~\cite{stieffenhofer_adversarial_2020} trained generative adversarial networks to reconstruct protein structural ensembles.
Similar tasks have been pursued with different architectures~\cite{yang_chemically_2023,jones_diamondback_2023}.
Our previous work seeks not only to add stochasticity to this task but to ensure that the sampled distribution can be reweighted to ensure Boltzmann statistics: we carry this out by training normalizing flows that conditionally sample rotamers given a backbone conformation~\cite{chennakesavalu_ensuring_2023}.
However, unlike Refs.~\cite{yang_chemically_2023, jones_diamondback_2023}, the models developed in Ref.~\cite{chennakesavalu_ensuring_2023} trained for individual proteins and are not general purpose.

\paragraph{Monte Carlo with generative models}
A number of recent works have sought to improve sampling using generative machine learning.
For Markov Chain Monte Carlo, \cite{noe_boltzmann_2019} used normalizing flows that were optimized to sample Boltzmann distributions, though this problem requires task-specific data~\cite{gabrie_adaptive_2022} and is difficult to scale to large systems.
Alternative approaches using distinct strategies for Monte Carlo~\cite{levy_generalizing_2018,song_-nice-mc_2017} have not found applications in physical systems due to complex training objectives.
Generative models have also been used to refine data and improve accuracy~\cite{maier_bypassing_2022}. 

\begin{figure}[h]
    \centering
    \includegraphics[width=\linewidth]{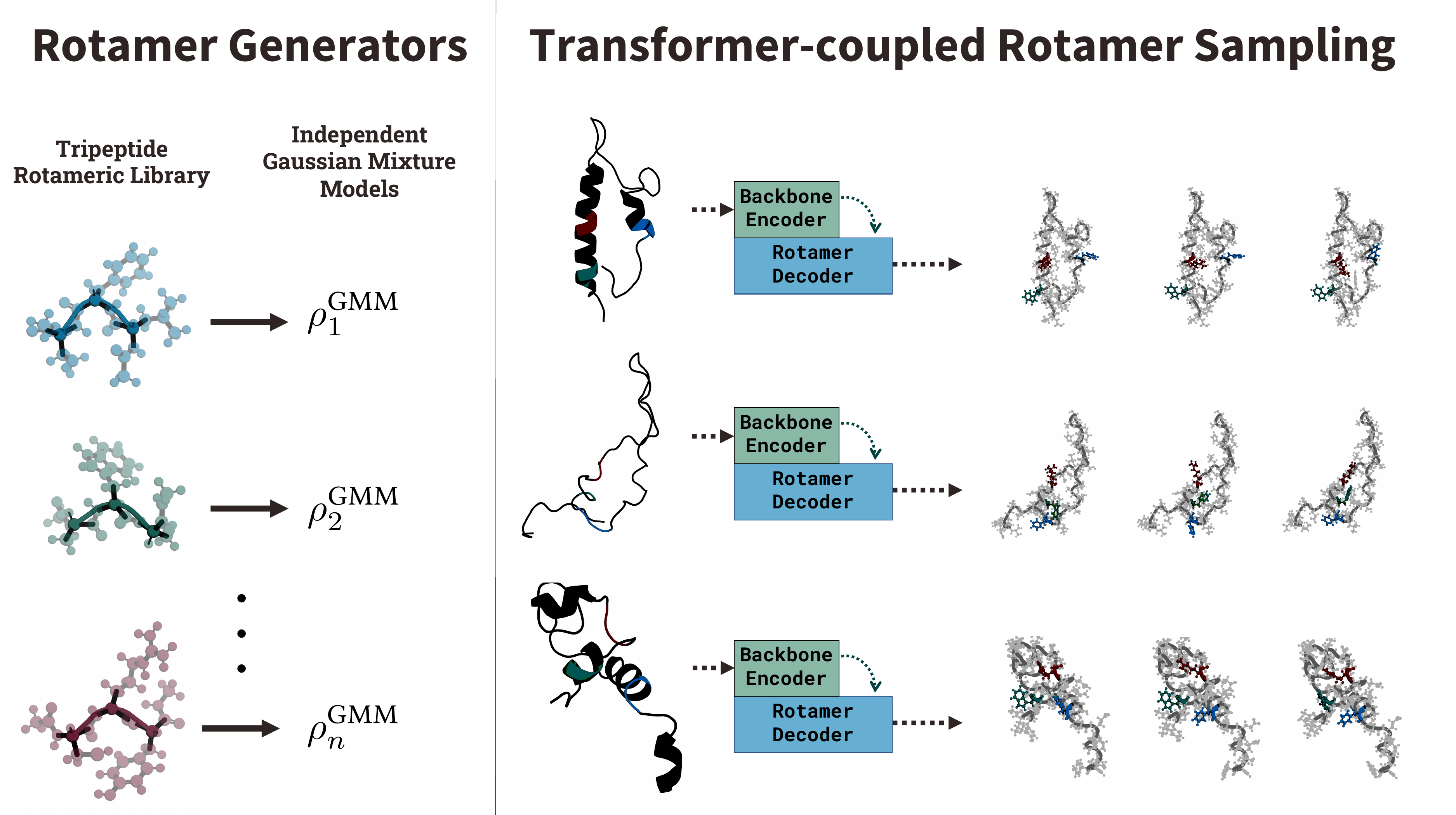}
    \caption{A schematic overview of backmapping procedure. First, for a given protein, a rotameric library of all unique tripeptide fragments is generated via long MD simulations in explicit solvent. Next, for each of these tripeptides, a unique and independent rotamer generator,  represented by a Gaussian mixture model, is trained. Finally, these independent rotamer generators are coupled via a transformer, and for a given backbone structure, the transformer generates an ensemble of side chain rotamer states. Select generated side chains are colored and correspond to the respective colored tripeptides depicted on the left.}
    \label{fig:schematic}
\end{figure}

\section{Methods}
\subsection{Independent Rotamer Generators}

It has long been observed the protein side chains occupy a small number of rotameric states \cite{chanhrasekaran_studies_1970}. These observations have led to the curation of so-called rotamer libraries which tabulate metastable conformations of certain side chain atoms, usually by recording their $\chi$ dihedral angles \cite{bhat_analysis_1979, ponder_tertiary_1987}. The distribution of allowed $\chi$ dihedral angles is dependent on the backbone conformation and on adjacent side chains, and this has led to rotamer libraries that incorporate varying degrees of information about the backbone, primary sequence, secondary structure, and protein type \cite{dunbrack_jr_bayesian_1997, chakrabarti_interrelationships_2001, bower_prediction_1997, dunbrack_rotamer_2002}. The information from these rotamer libraries has been exploited in many different settings, including protein structure prediction \cite{adams_phenix_2010}, protein-ligand docking \cite{mashiach_firedock_2008, moghadasi_impact_2015, watkins_sidechain_2016}, and backmapping from coarse-grained simulations \cite{jumper_rapid_2017}. 

Recent work by Dicks and Wales constructed a backbone-independent, sequence-dependent rotamer library for all possible tripeptides consisting of naturally occurring amino acids \cite{dicks_exploiting_2022}. Still, many rotamer libraries, including \cite{dicks_exploiting_2022} only record a collection of metastable states corresponding to free-energy minima.
Furthermore, these calculations are performed in implicit solvent. 
While practically useful in many settings, this information does not capture the full rotameric entropy of a side chain, which is crucial for statistically meaningful reconstruction of the side chain.

Inspired by the tripeptide simulations of \cite{dicks_exploiting_2022}, we sought to construct our own rotamer library that could capture the full rotameric entropy in a sequence-dependent fashion. To do so, we ran extensive molecular dynamics simulations of tripeptides in explicit solvent, and for every saved frame, recorded the appropriate internal coordinates for each atom in the side chain (see Appendix~\ref{app:tpp_sim} for full details). Given the relatively small size of these tripeptide systems, these simulations offer an appealing route for carrying out data acquisition for larger protein systems. A protein of interest can be fragmented into individual tripeptides, and through MD simulations, these tripeptide simulations can produce large amounts of data about rotamer states for individual side chains of the protein, avoiding the scaling bottlenecks of MD.

Many state-of-the-art generative models perform best in the large data limit, meaning that if we can leverage the training data of the generated library, we can use this data to build accurate generative models. For a given tripeptide, we train a model on side chain dihedral angles sampled during an MD simulation and use that model to then carry out rotamer sampling. Given that the number of dihedrals for any single side chain is relatively small \cite{chanhrasekaran_studies_1970} and that the distribution of the $\chi$ dihedrals generally has well-separated modes each of which appears to be approximately Gaussian (see Appendix~\ref{app:gmm}),  we use a Gaussian mixture model (GMM) to represent our individual rotamer generators. 
The form of the distribution for each residue is
\begin{equation}
    \rho^{\rm GMM} = \sum_{i = 1}^{K} \phi_i \mathcal{N}(\mu_i, \Sigma_i)
    \label{eq:gmm}
\end{equation}
where $K$ depends on the residue type and is chosen by inspection, though we note that this could also be done with a metric like the Bayesian information criterion. 

Using GMMs offers two advantages. First, training a GMM on with data from the generated rotamer library is computationally inexpensive taking only a few CPU minutes per residue. Generative models based on deep neural networks such as normalizing flows or diffusion models can require hundreds of GPU-hours~\cite{wirnsberger_normalizing_2021}. 
Second, each Gaussian component of the GMM can be loosely interpreted as a rotameric state of a particular side chain. While this interpretation is more tenuous for certain atoms in side chains like serine and threonine (see Figure~\ref{fig:app_dihedral_gmm}), where rotamer modes may be less Gaussian-like, it holds for most atoms across all side chains. In sum, the GMM-based rotamer generators enable rapid sampling of rotamer states for a given residue, providing a means of carrying out backmapping from a backbone configuration.

\subsection{Transforming backbone states to side chain distributions}
\label{sec:transformer}

A scheme that employs independent sampling of residue-wise rotamer generators will face a computational bottleneck; most generated structures will be statistically unlikely due to unfavorable energetics. Moreover, for large proteins, there are strong correlations among side chain rotamers, even for residues that are spatially and sequentially distant~\cite{dubay_long-range_2011}. This bottleneck can be avoided by coupling the independent generators, though this requires both learning correlations between side chains and conditioning on the backbone structure. A scheme that couples rotamer generators---which individually are faithful to the underlying statistics of the side chain---amounts to enforcing a strong physical prior on the backmapping process. 

While they are not determinative of a single backbone conformation, the $\phi$ and $\psi$ backbone dihedral angles provide information about secondary structure motifs \cite{ramachandran_stereochemistry_1963} and importantly constrain both the backbone and side chain dihedral angles \cite{chakrabarti_interrelationships_2001, harder_rotamers_2010}. Parameterizing the backbone by a sequence of $\phi$ and $\psi$ angles, we seek to generate a \textit{sequence} of rotameric states conditioned on the \textit{sequence} of backbone conformations. This requires a sequence-to-sequence transformation that captures both short-ranged and long-ranged correlations among components in the sequence. The multi-head self-attention mechanism \cite{vaswani_attention_2017}, foundational to many state-of-the-art seq2seq approaches, is well-suited to this task. 

The architecture we use builds on the standard encoder-decoder-based transformer architecture~\cite{vaswani_attention_2017}, where the backbone encoder block inputs a backbone sequence and the rotamer decoder block autoregressively generates the rotamer state. Our encoder and decoder blocks both use a multi-head self-attention mechanism with 8 heads, 6 layers, and an embedding dimension of 256. The generation process begins with a continuous backbone tokenization scheme in which the backbone coordinates $\xb^{(n)} \in \RR^{3n}$ are featurized as
\begin{equation}
    (\xb_1, \dots, \xb_n) \mapsto \begin{bmatrix}
\sin(\phi_1) & \sin(\phi_2) & \cdots & \sin(\phi_m) \\
\cos(\phi_1) & \cos(\phi_2) & \cdots & \cos(\phi_m) \\
\sin(\psi_1) & \sin(\psi_2) & \cdots & \sin(\psi_m) \\
\cos(\psi_1) & \cos(\psi_2) & \cdots & \cos(\psi_m)
\end{bmatrix},
\label{eq:bb_cont_token}
\end{equation}
where $m$ is the number of residues. We use the featurization in Eq.~\ref{eq:bb_cont_token} to avoid ambiguities with the periodicity of the dihedral angle coordinates. We also incorporate a categorical backbone tokenization scheme that embeds the residue identity using a standard lookup-table-based learnable embedding layer. The continuous and categorical embeddings are concatenated before being passed through a standard sinusoidal positional encoding layer. This embedded backbone sequence is then passed into the encoder block of the transformer.

Relying on the interpretation that each Gaussian component of a rotamer generator corresponds to a particular rotamer state, we autoregressively output a sequence of Gaussian components via the rotamer decoder block (see Figure~\ref{fig:schematic}). Ideally, the transformer would generate a sequence of side chain rotamer states that are statistically likely under a thermal Boltzmann distribution, though ensuring this will require additional algorithmic developments as discussed in Sec~\ref{sec:outlook}.  Upon determining a sequence of Gaussian components, we can then straightforwardly generate the coordinates of all side chain atoms by sampling internal coordinates from the selected Gaussian components (see Appendix~\ref{app:bmap} for further details). We emphasize that the side chains that are generated are correlated through the attention mechanism of the transformer decoder block. While this procedure enables correlations that span the entire protein chain to develop, we believe that incorporating energetic information will aid generalization~\cite{chennakesavalu_ensuring_2023, vigueradiez_generation_2023}. For the systems we study here, the training data is collected by sampling rotamer generators independently, carrying out a short relaxation step, and retaining all configurations below a fixed cutoff energy. We note that this initial data collection is meant primarily to provide a test of the approach; in a more general setting, we envision relying on additional and diverse protein structural and trajectory data (see Section~\ref{sec:outlook}). 
We compare our method to other machine learning based backmapping methods in Appendix~\ref{app:benchmark} and find that our technique is competitive on all metrics and superior in some, showing high diversity in the generated conformations while maintaining a very low propensity for steric clashes.

\section{Results}
\begin{figure}
    \centering
    \includegraphics[width=1.0\linewidth]{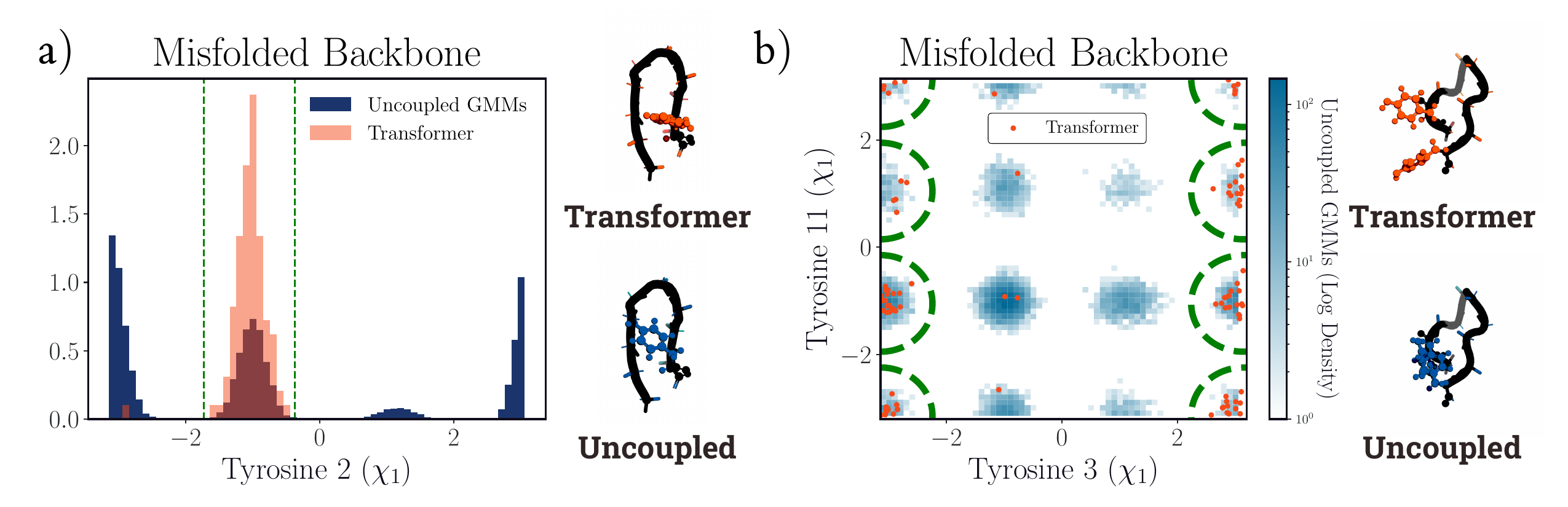}
    \caption{Backmapping generates physically meaningful rotamer states of terminal bulky side chains for a representative misfolded backbone. While all side chains are backmapped, only select residues are shown for clarity. a) Uncoupled sampling of Tyr2 results in samples with backbone clashes, while transformer-coupled sampling restricts sampling to physically allowed mode delineated in green (centered at $-\frac{\pi}{3}$). b) Transformer-coupled sampling accurately captures correlations between Tyr3 and Tyr11 and restricts sampling to allowed modes delineated in green, while uncoupled sampling results in samples with steric clashes.}
    \label{fig:chigtyr}
\end{figure}

\subsection{Chignolin}
\label{sec:chig_result}
The miniprotein Chignolin is a fast-folding, ten-residue protein. At equilibrium, Chignolin fluctuates between three metastable states: an unfolded state, a folded state, and a misfolded state \cite{lindorff-larsen_how_2011, husic_coarse_2020}. Here, we consider the CLN205 variant of Chignolin which is known to adopt a $\beta-$hairpin structure in its folded state \cite{honda_crystal_2008a}. The mechanisms of Chignolin folding are well-studied and occur first via a ``hydrophobic collapse,'' followed by further rearrangements \cite{mckiernan_modeling_2017a}. Here, we assume that we have access to a collection of coarse-grained backbone configurations that we are interested in backmapping from, which we collect by coarse-graining atomistic configurations from a long all-atom MD simulation \cite{husic_coarse_2020}. 

Chignolin can be split into 8 unique tripeptides, and for each of these tripeptides, we carry out long MD simulations in explicit solvent. To simplify the generation process, we work with internal coordinates, namely bond lengths, bond angles, and dihedral angles, and for each side chain atom, we record the appropriate set of internal coordinates in our generated rotamer library. For each residue, using a standard Expectation-Maximization (EM) algorithm, we train two Gaussian mixture models (GMMs): a GMM trained on the joint distribution of bond lengths and bond angles and a second GMM trained on the dihedral angles (See Appendix~\ref{app:gmm} for further details). These rotamer generators can then be used to iteratively reconstruct an entire side chain. 

A backmapping strategy that can approximately sample the true underlying conditional Boltzmann distribution of course requires a way to couple these independent rotamer generators, which we do using a transformer (see Section~\ref{sec:transformer}, Appendix~\ref{app:trans} for further details). With Chignolin, for each of the 50000 backbone configurations, we generate 10000 possible rotamer states by independently sampling all rotamer generators, relax those structures using a short MD step, and select all structures with energies below a threshold (see Appendix~\ref{app:prot}). We emphasize that this dataset does \textit{not} consist of Boltzmann-distributed samples, but it does consist of configurations that individually have a reasonable probability mass within the Boltzmann distribution. Finally, we use an 80/10/10 train/validation/test-set split, where we carry out the split across backbone configurations, so no single backbone configuration is present in two or more of these split datasets. We train the transformer using a standard cross-entropy loss, select the best-performing network on the validation set, and carry out all analyses below on the test set.

\subsubsection*{Backmapping of terminal side chains involved in hydrophobic collapse}
The terminal bulky residues of Chignolin are primarily implicated in hydrophobic collapse \cite{mckiernan_modeling_2017a}, and first, we investigate our ability to backmap these residues. The misfolded state of Chignolin is characterized by the N-terminal end of the protein bending towards the interior of the hairpin. Practically, this results in an interior that is more tightly packed than the folded or unfolded state. As an initial test, we consider the distribution of $\chi_{1}$ dihedral angles for the first tyrosine residue (Tyr2) in the sequence on a representative misfolded backbone. The $\chi_{1}$ dihedral angle influences the position of the $\gamma-$Carbon and can be intuited as the overall orientation of the residue relative to the backbone, with downstream dihedral angles $\chi_{2\dots K}$ influencing the local structure of the Tyrosine ring.

In Figure~\ref{fig:chigtyr}, we plot the distribution of $\chi_1$ generated via uncoupled rotamer generators and compare with $\chi_1$ generated from transformer-coupled rotamer sampling. First, we note that during sampling from the independent rotamer generators, we sample from the three different rotamer states (centered at -$\frac{\pi}{3}$, $\frac{\pi}{3}$, and $\pi$) in a proportion commensurate with that observed during the respective tripeptide simulation. For the representative misfolded backbone, the rotamer state centered at $\pi$ will result in Tyr2 sterically clashing with the backbone, while the state centered at $\pi/3$ is a statistically likely higher energy rotamer state. With transformer-informed sampling, sampling is limited to a statistically likely rotamer state.

\begin{figure}[t]
    \centering
    \includegraphics[width=\linewidth]{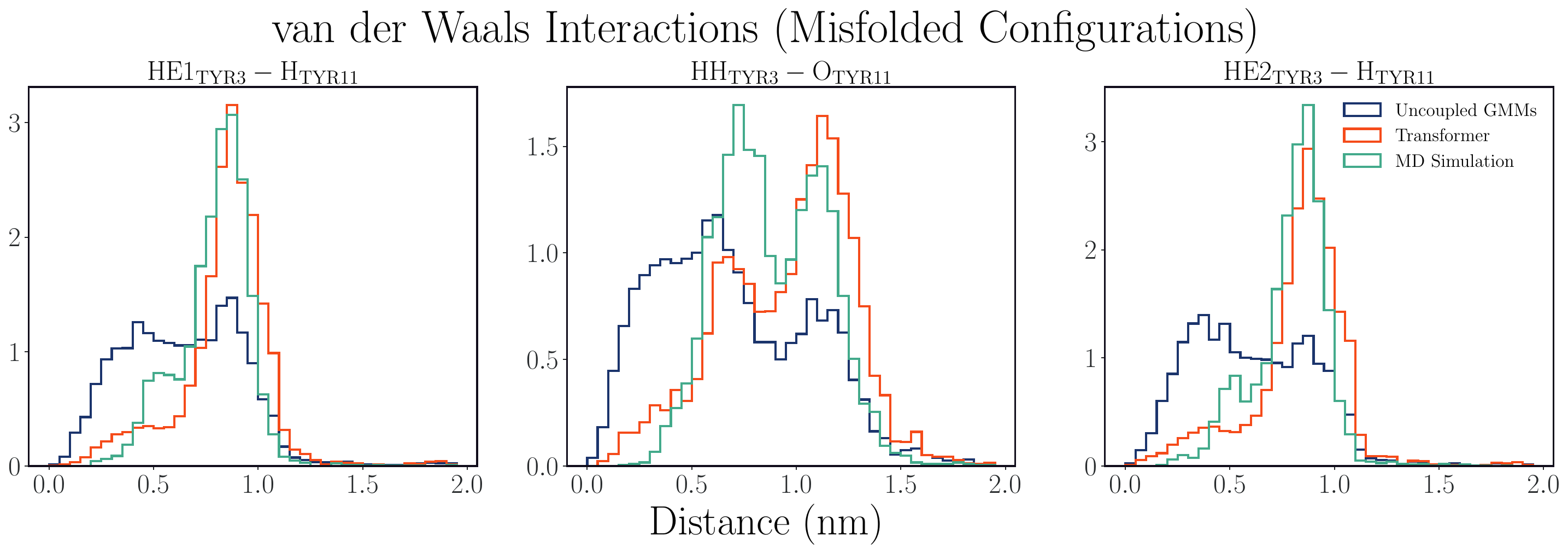}
    \caption{Backmapping captures van der Waals interactions between nearby non-bonded atoms for an ensemble of misfolded backbones. For select pairs of atoms highlighted, uncoupled sampling results in samples with high-energy interatomic distances; with transformer-coupled sampling, unphysical distances are greatly minimized, and importantly, are in strong agreement with distances observed during long MD simulations.}
    \label{fig:chig_dist_close}
\end{figure}

Next, we consider the correlations between Tyrosine 3 and Tyrosine 11, two additional bulky side chains also involved in hydrophobic collapse. While spatially proximal in a collapsed state, these residues are sequentially distant, providing an important test for whether the transformer-coupled sampling captures long-range correlations in the backbone sequence. In Figure~\ref{fig:chigtyr}, we plot the distribution of the respective distribution of $\chi_1$ dihedral angles for the two residues. With uncoupled sampling, all 9 rotamer states are sampled to differing proportions; however, with transformer-coupled sampling, most of the samples are within the 3 energetically favorable modes---delineated in green---with only a few samples in modes that are statistically unlikely either as a result of a side-chain-backbone clash or a side-chain-side-chain clash. 

Finally, we plot the distribution of all rotamer states across all backbones in Appendix~\ref{app:prot_chig}. We observe that as the hairpin interior becomes more obstructed, the rotamer entropy decreases; more states become disallowed. This is a trend that we also observe in the MD simulation \cite{husic_coarse_2020}. We note that the MD simulation is not exhaustive and does not fully sample all rotamer states and so only provides qualitative insight into the effect of different backbone metastable states on the rotamer distribution.

\begin{figure}
    \centering
    \includegraphics[width=0.4\linewidth]{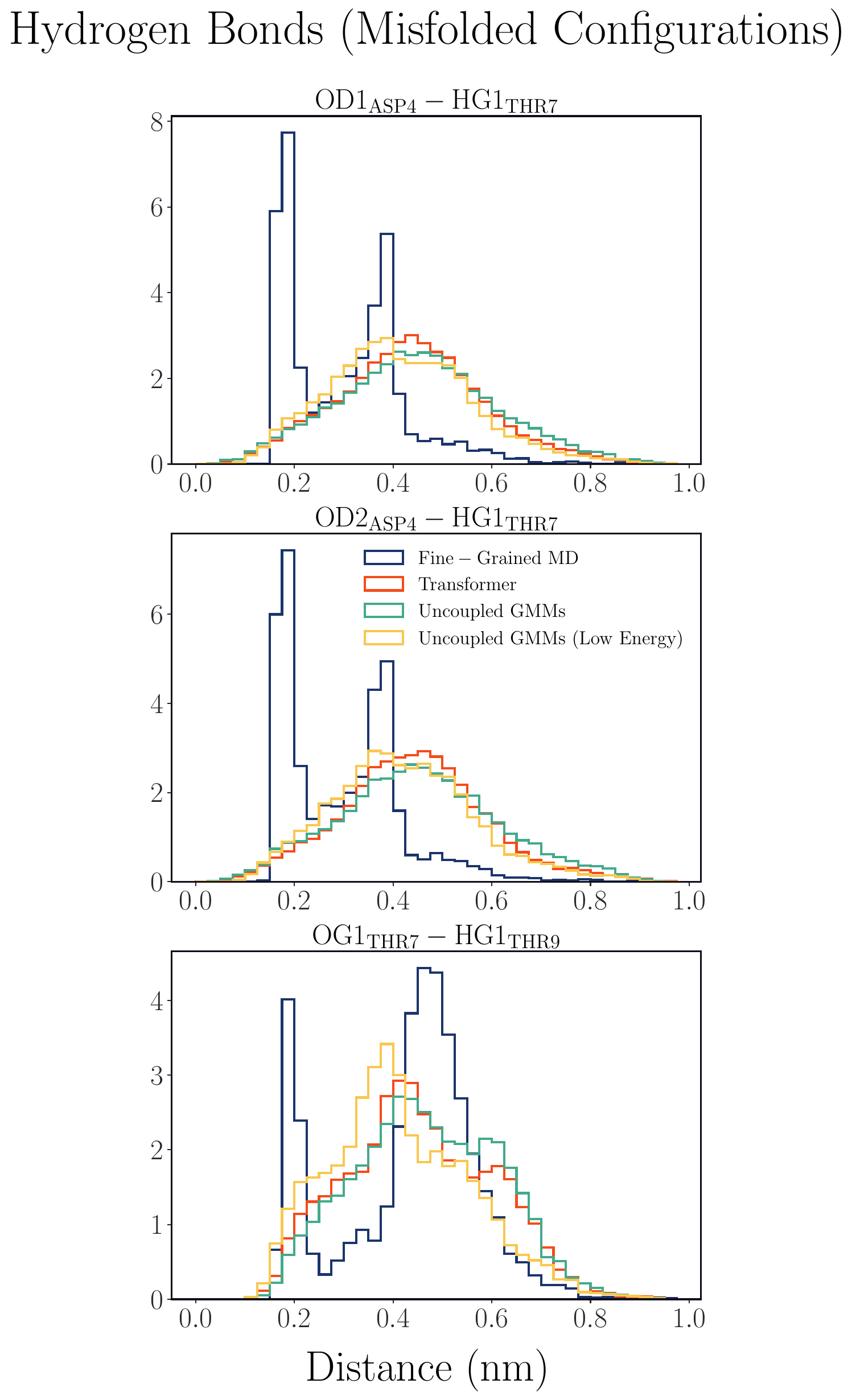}
    \caption{Backmapping does not fully capture hydrogen bonds for an ensemble of misfolded backbones. For atoms that engage in hydrogen bonds, transformer-coupled sampling does not in interatomic distances in agreement with long MD simulation. This is primarily a limit of the data curation strategy as evidenced by the agreement with relaxed, low-energy samples (yellow). Strategies for overcoming this are outlined in Section~\ref{sec:outlook}.}
    \label{fig:chig_h_bond}
\end{figure}

\subsubsection*{Analysis of non-bonded interactions}

The backmapping strategy we employ here relies on strong implicit physical priors about allowed bond lengths, bond angles, and dihedrals through individual rotamer generators. However, it does not incorporate priors about non-bonded distances.
To determine the efficacy of our approach, we examine select non-bonded interactions of Chignolin. First, we consider pairs of side chain atoms that have only non-bonded van der Waals and electrostatic interactions. We plot the distribution of distances between three of these pairs in Figure~\ref{fig:chig_dist_close}, which consists of various atom pairs between Tyrosine 3 and Tyrosine 11. We observe that uncoupled sampling can often generate configurations with non-bonded distances that are unphysical. With transformer-coupled sampling, we minimize the generation of these distances and generate configurations with distances that are near-commensurate with those observed in an MD simulation. 

Second, we consider pairs of side chain atoms that engage in hydrogen bonding. There are three persistent side-chain-side-chain hydrogen bonds in Chignolin, which occur between various atoms in Thr7, Thr9, and Asp4. We plot the distributions of these distances in Figure~\ref{fig:chig_h_bond}. We observe that the transformer-coupled sampling does not generate configurations with hydrogen bonding comparable to the hydrogen bonding present in the MD dataset. This likely results from our dataset curation strategy, where the training dataset does not consist of Boltzmann-distributed configurations but rather only those that are below some energy threshold. This is also supported in Figure~\ref{fig:chig_h_bond}, where we also plot the distribution of distances of low energy configurations from relaxed uncoupled sampling. Ultimately, this highlights the limitation of using a purely ``bottom-up'' strategy to generate a training dataset, which fails to appropriately weigh relevant phenomena like hydrogen bonding. We provide strategies for carrying out reweighting to recapitulate physically important phenomena like hydrogen bonding in Sec.~\ref{sec:outlook}.

\subsection{Intrinsically Disordered Region of the Androgen Receptor (AR-IDR)}
\begin{figure}
    \centering
    \includegraphics[width=\linewidth]{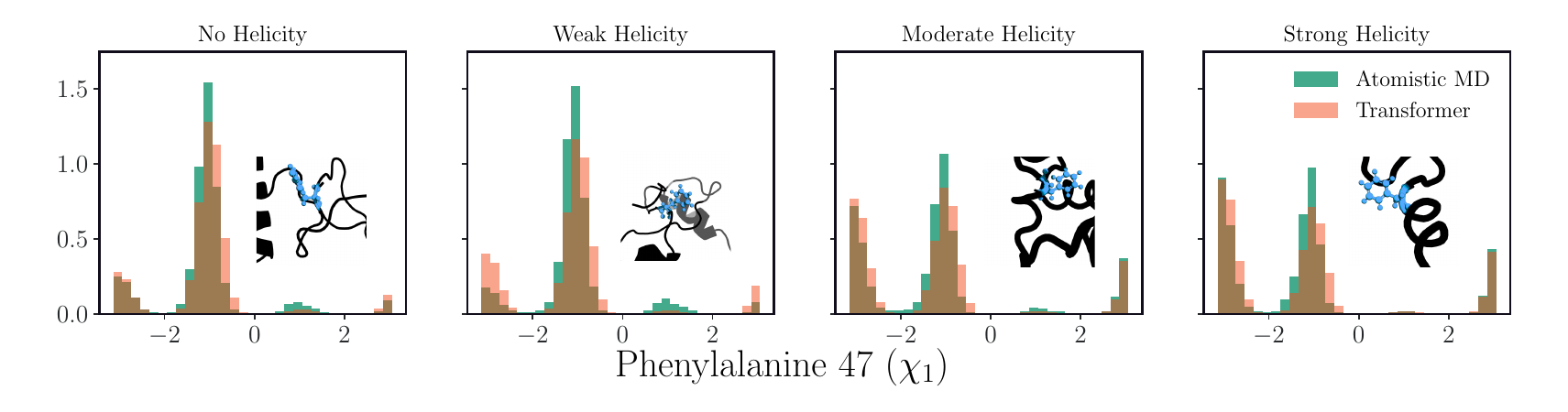}
    \caption{Transformer-coupled sampling accurately captures the effect of backbone helicity on the rotamer state of Phe47. As the degree of helicity increases, the mode centered at $\pi$ becomes more populated, while the mode centered at $-\frac{\pi}{3}$ becomes less populated. In the insets, representative configurations are plotted, with the left two plots consisting of samples from the mode centered at $-\frac{\pi}{3}$, with the right two plots consisting of samples from the mode centered at $\pi$.}
    \label{fig:helicity}
\end{figure}
Intrinsically disordered proteins (IDPs) are a class of proteins that adopt an ensemble of heterogeneous configurations, with transient local secondary structure propensity \cite{dyson_intrinsically_2005, wright_intrinsically_2015}. Gaining mechanistic insight into IDP function has tremendous scientific value due to role of these proteins in an enormous array of biological processes and phenomena including transcriptional regulation \cite{zhu_small_2022a}, cellular signal transduction \cite{wright_intrinsically_1999, wright_intrinsically_2015}, and membrane-less organelles \cite{uversky_intrinsically_2015, brangwynne_polymer_2015}. 
Characterizing the conformational ensemble with atomistic detail constitutes an important first step to piecing apart the structure-ensemble-function relationship for IDPs, though this is not feasible experimentally with NMR or SAXS \cite{bonomi_principles_2017} or computational techniques AlphaFold---\cite{jumper_highly_2021, ruff_alphafold_2021} which ultimately provides only a small number of structures of questionable thermodynamic relevance. 
All-atom and coarse-grained simulations offer a promising route towards high-resolution characterization of the structural ensemble, and in light of this promise, extensive effort has been dedicated to developing atomistic \cite{robustelli_developing_2018} and coarse-grained force fields \cite{vitalis_absinth_2009, latham_maximum_2020, latham_unifying_2022}.  

Coarse-grained simulations paired with backmapping can provide detailed insight into the structural ensemble in a computationally tractable fashion. However, carrying out backmapping from coarse-grained trajectories of IDPs is an especially challenging task as a robust strategy for backmapping will have to contend with sampling rotameric states across a diversity of backbone structures. Here, we investigate our ability to sample rotameric states for an intrinsically disordered region of the androgen receptor (AR) \cite{demol_epi001_2016, zhu_small_2022a}. AR is a transcription factor that contains folded DNA and ligand-binding domains and an intrinsically disordered N-terminal transactivation domain (NTD) \cite{reid_conformational_2002, lavery_structural_2008}. Recent work by Zhu et al.\ investigated a particular 56 residue region of the NTD, which is bound by two small-molecule inhibitors; using all-atom simulations, they observed that these inhibitors induce increased helicity in the disordered region, restricting transcription activation \cite{zhu_small_2022a}. Here, we investigate our ability to carry out backmapping of this intrinsically disordered region, and for notational simplicity, we henceforth term this region AR-IDR.

As in Section~\ref{sec:chig_result}, we assume that we have access to a collection of coarse-grained configurations, which we generate by coarse-graining atomistic configurations from a long all-atom MD simulation \cite{zhu_small_2022a}. AR-IDR has 51 unique tripeptides, and we similarly carry out long MD simulations for each of these tripeptides in explicit solvent and train corresponding rotamer generators. With AR-IDR, we had access to 57144 backbone configurations, and for each of these configurations, we generated 50000 possible rotamer states by independently sampling all rotamer generators, carried out a short relaxation step, and selected structures with energies below a threshold (see Appendix~\ref{app:prot}). Using again an 80/10/10 train/validation/test-set split, we select the best-performing network on the validation set and carry out all analyses below on the test set. 

\subsubsection*{Backbone helicity informs rotameric sampling}

In the apo state, the AR-IDR can adopt local helical structures (see Figure~\ref{fig:schematic} and Figure~\ref{fig:helicity} inset), and the degree of helicity influences the allowed rotamer states. Furthermore, two small-molecule inhibitors that target AR-NTD interact with various aromatic residues (Trp7, Trp43, Phe47) \cite{zhu_small_2022a} that are present in regions that have high helical propensity. As a first test, we examined whether our backmapping strategy could capture the rotamer density of these residues across differing degrees of helicity. 

In Figure~\ref{fig:helicity}, we consider the effect of helicity on Phe47. Using an order parameter that measures discrepancies from an ideal helix, we categorize the backbone region adjacent to Phe47 into four categories: no helicity, weak helicity, moderate helicity, and strong helicity. We then investigate the effect of this backbone structure on the $\chi_1$ dihedral angle of Phe47 and compare it to the rotameric states observed during MD simulation. We find that under transformer-coupled sampling, as the degree of helicity increases, there is increased occupancy of the rotamer state centered at $-\pi$ and decreased occupancy of the rotamer state centered at $-\frac{\pi}{3}$. Remarkably, this is a trend that we also observe from atomistic MD simulations, demonstrating that we can capture this effect without access to an existing dataset of rotameric states for this protein. We analyze the consequence of helicity for Trp7 and Trp43 in Appendix~\ref{app:prot_ar}.

\begin{figure}
    \centering
    \includegraphics[width=0.7\linewidth]{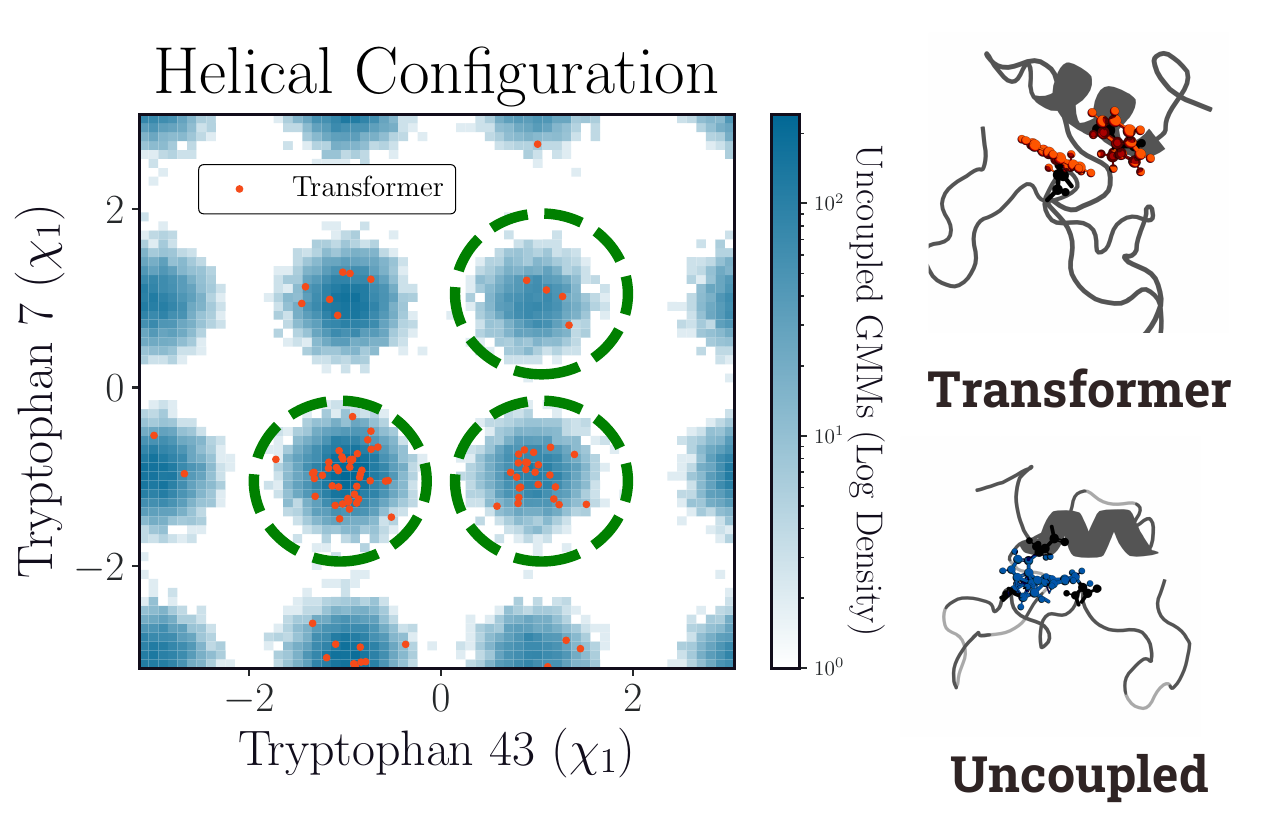}
    \caption{Transformer-coupled sampling accurately captures sequentially long-range correlations between Trp7 and Trp43 for a representative helical molten-globule state. The three modes centered at $(-\frac{\pi}{3}, -\frac{\pi}{3})$, $(\frac{\pi}{3}, -\frac{\pi}{3})$, and $(\frac{\pi}{3}, \frac{\pi}{3})$ correspond to the lowest energy metastable rotamer states and transformer-coupled sampling generally restricts sampling to these modes.}
\label{fig:ar_coupling}
\end{figure}

\subsubsection*{Correlated backmapping of aromatic side chains}

Next, we look at a helical ``molten-globule''-like state, representative of the structure that AR-IDR maintains when bound by a small-molecule inhibitor. This more compact nature of the protein is a more difficult setting to carry backmapping in due to the close spatial proximity of many residues distant in the primary sequence. In Figure~\ref{fig:ar_coupling}, we consider a backbone conformation where Trp7 and Trp43 are spatially close, requiring a transformer-based sampling scheme to couple sampling of the respective rotamer states.

For this backbone state, the three modes centered at $(-\frac{\pi}{3}, -\frac{\pi}{3})$, $(\frac{\pi}{3}, -\frac{\pi}{3})$, and $(\frac{\pi}{3}, \frac{\pi}{3})$ correspond to the lowest energy metastable rotamer states, while the modes centered at $(-\frac{\pi}{3}, \pi)$,  $(\frac{\pi}{3}, \pi)$, and $(\pi, \pi),$ would result in rotamer states with unphysical steric overlaps. Finally, the remaining three modes are less favorable statistically but do not generally correspond to rotamer states with steric overlaps. From Figure~\ref{fig:ar_coupling}, it is clear that transformer-informed sampling directs sampling towards statistically favorable modes (outlined in green) while restricting sampling from statistically unfavorable and unlikely modes. Finally, we note that the imperfections in sampling can be corrected with a reweighting scheme. While this is not outright possible with transformer-informed sampling, we outline strategies for reweighting in Section~\ref{sec:outlook}.

\section{Discussion \& Conclusions}
\label{sec:outlook}
Generative models offer tremendous promise for rapidly sampling large biomolecular systems; however, relying on such a strategy is only viable if we can deploy generative models in settings with task-limited data. In this paper, we outline an approach to sample the rotameric density of proteins via a modular backmapping approach. Crucially, this approach requires no data from the target density \textit{a priori}, and instead, we leverage cheap-to-acquire data from related systems, namely tripeptide simulations, to carry out backmapping.

While this data-free, transformer-based sampling scheme is promising, one of the drawbacks, as discussed earlier, is our inability to reweight samples and provide guarantees that the sampling is Boltzmann distributed. Currently, the model we have proposed does not directly allow for Boltzmann reweighting because 1) we do not know the probability distribution over the backbone structures and 2) the transformer architecture that we use is not directly invertible. 
In previous work, we carried out Boltzmann sampling by building a coarse-grained model for the backbone and subsequently backmapping with an invertible normalizing flow model~\cite{chennakesavalu_ensuring_2023}. Normalizing flows are a class of generative models with bespoke architectures designed to enable exact likelihood computation, that, like many neural networks, perform remarkably well in the large-data limit. One feasible approach to carry out reweighting would be to generate large amounts of data using the transformer, and with this data, train a normalizing flow to carry out backmapping. Because the transformer generates samples that have reasonable likelihood under the Boltzmann distribution, reweighting will be more successful than a normalizing flow directly trained on samples from uncoupled rotamer generators. 
This strategy would also provide a greater diversity of configurations than atomistic MD and would not be susceptible to mode collapse~\cite{gabrie_adaptive_2022}. Alternatively, it may be possible to formulate a sampling strategy in the pseudo-marginal MCMC framework~\cite{andrieu_pseudo-marginal_2009}; however, this requires additional theoretical development. 

Integrating a diverse set of data streams will be essential if we are to build scalable generative models that accurately reflect the thermal fluctuations of complex biomolecules. 
While the approach we outline in this work is ``bottom-up'' in the sense that physics-based simulations of tripeptides provide models that are then adapted to specific systems, we also foresee employing a ``top-down'' approach in which the rotamer transformer is also trained on experimental structures.
The limitations we see in our present model we believe arise primarily from insufficient data, which a top-down approach will help alleviate. 
We expect that integrating this data will help capture a diversity of correlations among side chains in many distinct structural motifs. 

\section*{Acknowledgements}
The authors thank Aaron Dinner and David Wales for the suggestion of using rotamer libraries. The authors would also like to acknowledge Frank Hu for helpful discussions about transformers. This material is based upon work supported by the U.S. Department of Energy, Office of Science, Office of Basic Energy Sciences, under Award Number DE-SC0022917.

\paragraph*{Data and Code Availability:}
The data that support the findings of this study are available from the corresponding author upon reasonable request. Our code is available on GitHub at the following url: \href{https://github.com/rotskoff-group/transformer-backmapping}{https://github.com/rotskoff-group/transformer-backmapping}.

\paragraph*{Supporting information:}
Details for rotamer library, transformer, protein simulation and benchmarking against existing backmapping methods.


\printbibliography
\newpage 
\begin{figure}
    \centering
    \includegraphics[width=\linewidth]{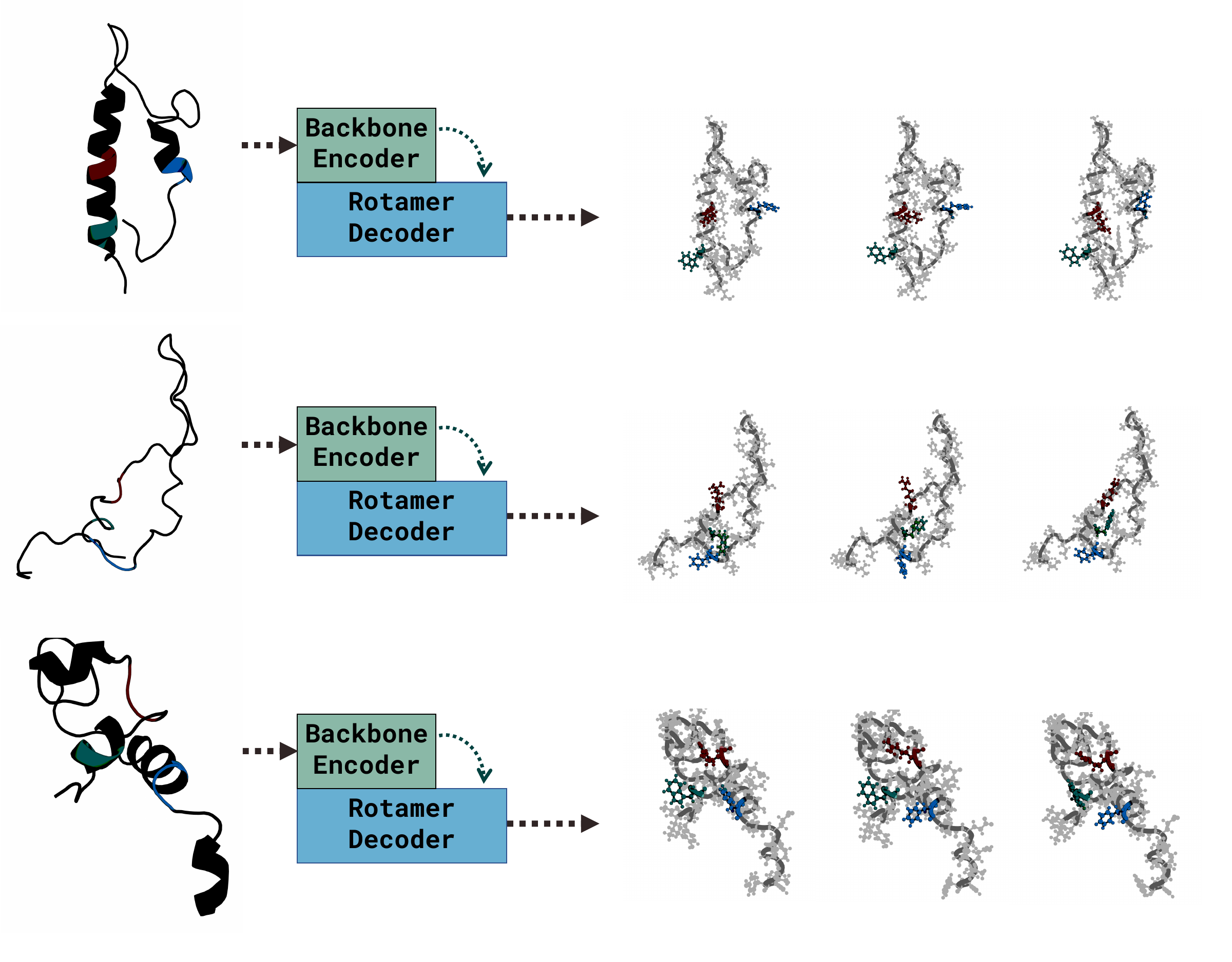}
    \caption*{TOC Graphic}
\end{figure}

\appendix
\setcounter{page}{1}    
\setcounter{figure}{0}  


\renewcommand\thefigure{S\arabic{figure}} 
\renewcommand\thepage{S\arabic{page}}     
\clearpage
\newpage


\section{Rotamer Library}
\subsection{Tripeptide simulations}
\label{app:tpp_sim}

We ran MD simulations of tripeptides in explicit solvent to exhaustively sample the rotameric density. Each simulation was conducted in the NVT ensemble using the under-damped BAOAB integrator~\cite{leimkuhler_rational_2012} with a time step of 0.001 $\rm{ps}$ and a friction coefficient of $\gamma = 0.1 $ $\rm{ps}^{-1}$. We did not incorporate any bond constraints. While treating bonds as constraints allows a larger integration time step, flexible bonds better capture the fluctuations of the tripeptide, which is desirable for reweighting. We ran each tripeptide simulation for 48 GPU hours at 300K, resulting in trajectories that ranged from 0.5-1 $\mu\textrm{s}.$

For simplicity, we only ran simulations for tripeptides that were relevant to Chignolin and AR-IDR. For each non-terminal residue $\textrm{res}_i$ in the protein, we ran a corresponding simulation for the tripeptide $\textrm{res}_{i-1}\textrm{res}_i\textrm{res}_{i+1},$ where we used information from the central residue for training the rotamer generator for residue $i$. For the first residue $\textrm{res}_1$ the corresponding tripeptide simulation was $\textrm{res}_1\textrm{res}_{2}\textrm{res}_{3}$ where we used information from the first residue of the tripeptide for training the rotamer generator. For the last residue $\textrm{res}_m,$ the corresponding tripeptide was $\textrm{res}_{m-2}\textrm{res}_{m-1}\textrm{res}_{m},$ where we used information from the last residue of the tripeptide for training the rotamer generators.

For Chignolin, there were 8 unique tripeptides and for AR-IDR, there were 51 unique tripeptides. For the Chignolin tripeptides we used the Amber ff14SB force field \cite{maier_ff14sb_2015} with the TIP3P water model \cite{jorgensen_comparison_1983}. For the AR-IDR tripeptides, we used the a99SB-disp force field and a99SB-disp water model \cite{robustelli_developing_2018}.

\subsection{Reconstructing a side chain}
\label{app:bmap}
We use an internal coordinate representation based on bond distances, bond angles, and dihedral angles to carry out side chain reconstruction. This representation is often collated in a ``$z$-matrix,'' where a sample $z$-matrix is, for example,
\begin{equation}
    \begin{bmatrix}
\textrm{atom}_1 & \textrm{atom}_2 & \textrm{atom}_3 & \textrm{atom}_4 \\
\textrm{atom}_2 & \textrm{atom}_3 & \textrm{atom}_4 & \textrm{atom}_5 \\
\textrm{atom}_3 & \textrm{atom}_4 & \textrm{atom}_5 & \textrm{atom}_6
\end{bmatrix}.
\end{equation}
Given the Cartesian coordinates of $\textrm{atom}_1,$ $\textrm{atom}_2,$ and $\textrm{atom}_3,$ the bond length between $\textrm{atom}_3$ and $\textrm{atom}_4,$ the bond angle between $\textrm{atom}_2, $$\textrm{atom}_3,$ and $\textrm{atom}_4,$ and the dihedral angle between  $\textrm{atom}_1,$ $\textrm{atom}_2,$ $\textrm{atom}_3,$ and $\textrm{atom}_4,$ it is straightforward to compute the Cartesian coordinate of $\textrm{atom}_4$ using elementary linear algebra. 

The appropriate coordinates, bond lengths, bond angles, and dihedrals can then be used to iteratively determine the coordinates of $\textrm{atom}_5,$ and $\textrm{atom}_6.$ For every residue considered in our investigation, we have defined a corresponding $z$-matrix, where the dihedral angle defined by the $i$th row corresponds to the dihedral angle $\chi_i.$ We refer the reader to our code for the definitions of a $z$-matrix for a particular residue.

From the tripeptide simulations (see Appendix~\ref{app:tpp_sim}), we recorded the appropriate bond lengths, bond angles, and dihedral angles for every row in our $z$-matrix, which we then used to train our rotamer generators.

\begin{table*}
\centering
\begin{tabular}{|p{2.5cm}|p{1cm}|}
 \hline
 \multicolumn{2}{|c|}{Number of Gaussian Components} \\
 \hline
 Residue Name &K\\
 \hline
 ACE & 3\\
 NME & 3\\
 ALA & 3\\
 ARG & 48\\
 ASN & 16\\
 ASP & 64\\
 CYS & 8\\
 GLN & 14\\
 GLU & 64\\
 GLY & N/A\\
 HIS & 8\\
 LEU & 64\\
 LYS & 96\\
 MET & 64\\
 PHE & 8\\
 PRO & 2\\
 SER & 12\\
 THR & 32\\
 TRP & 8\\
 TYR & 16\\
 VAL & 32\\
 \hline
\end{tabular}
\caption{Number of Gaussian components used for side chain dihedral generator. Residues not included here were not present in proteins investigated in this work.}
\label{tab:n_comp}
\end{table*}

\begin{figure}
    \centering
    \includegraphics[width=0.5\linewidth]{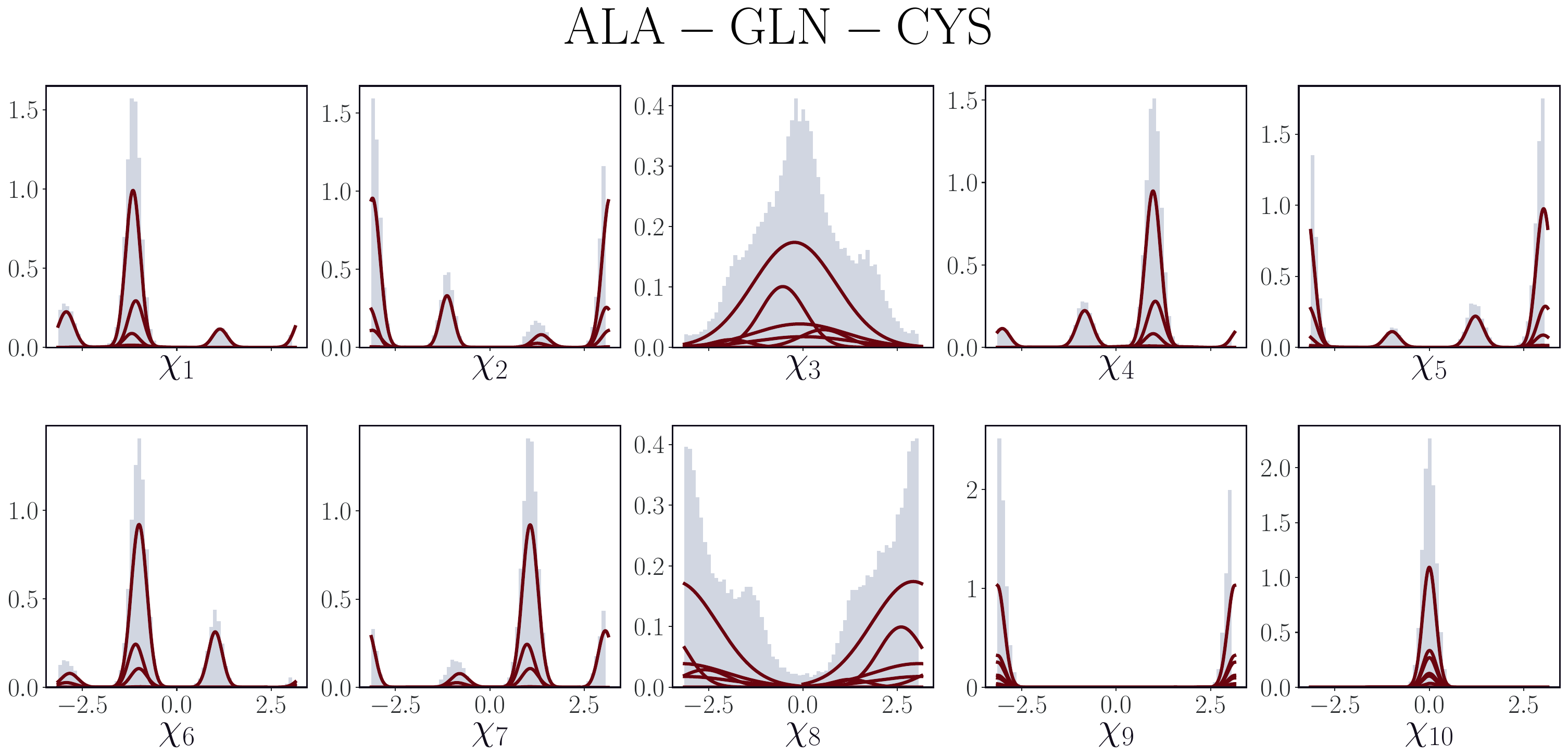}
    \includegraphics[width=0.5\linewidth]{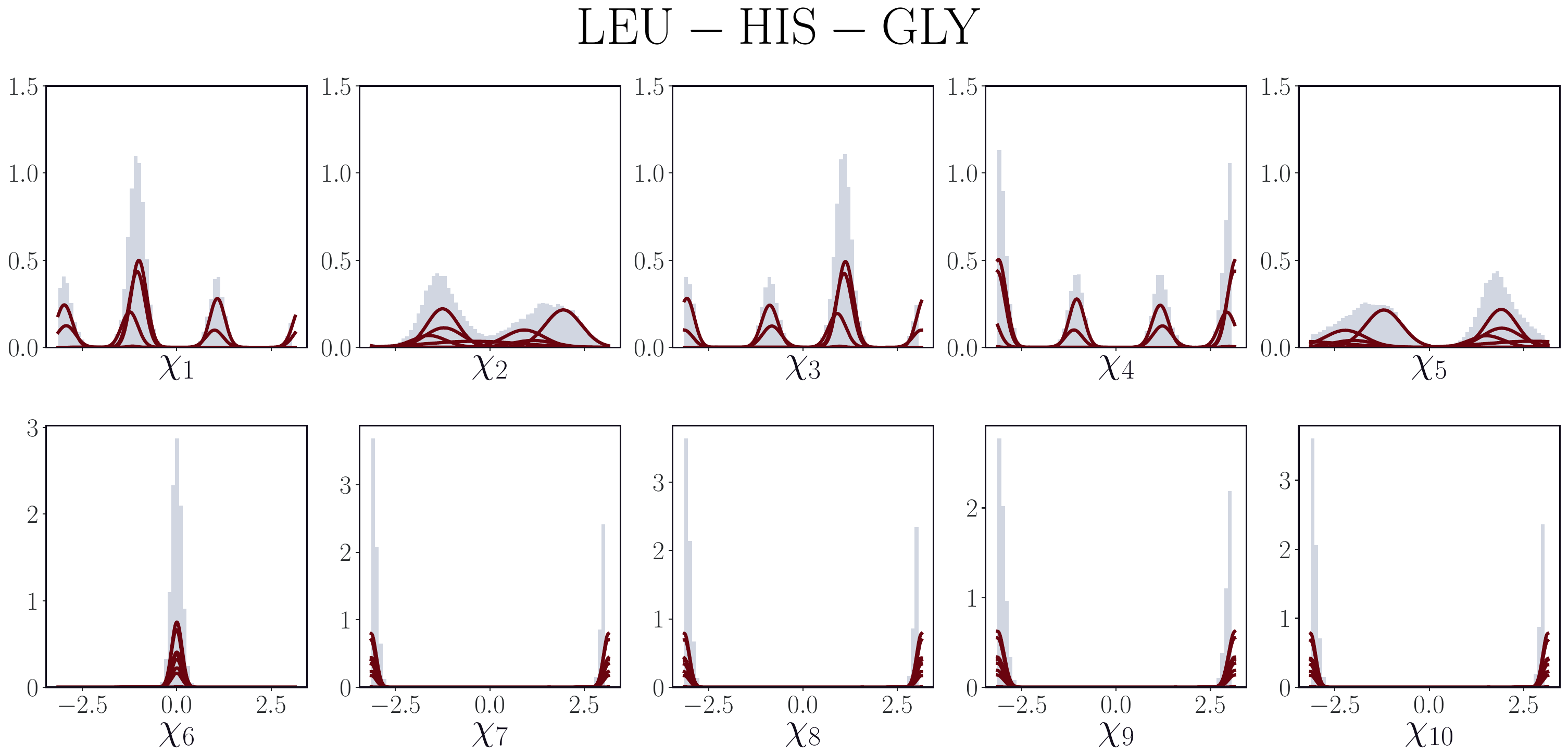}
    \includegraphics[width=0.5\linewidth]{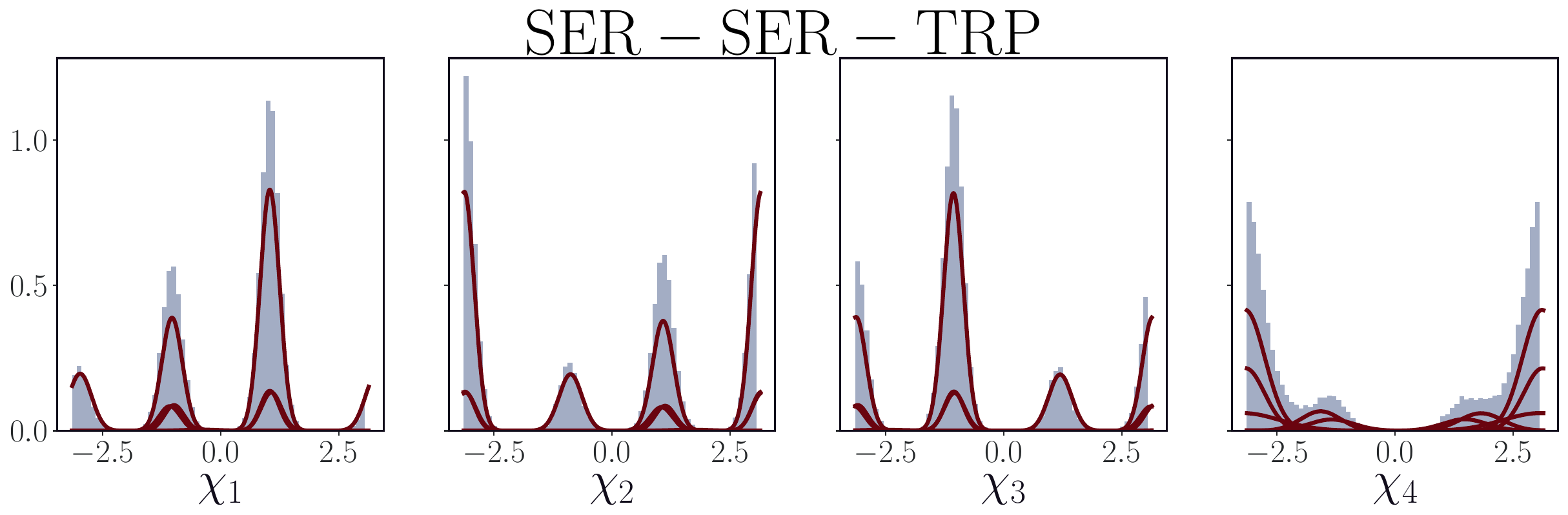}
    \caption{Distribution of $\chi$ dihedrals for select AR-IDR triptpide simulations. Components Gaussians are overlaid in maroon and are scaled by their mixture weight $\phi_i$. Most modes are Gaussian-like but certain modes (e.g.\ $\chi_{8}$ of GLN) are less Gaussian-like.}
    \label{fig:app_dihedral_gmm}
\end{figure}

\subsection{Gaussian Mixture Models}
\label{app:gmm}
We use a Gaussian Mixture Model (GMM) as a generator for the internal coordinates for a given residue.
In this representation, the distribution of $z$-matrix coordinates is approximated as,
\begin{equation}
    \rho^{\rm GMM} = \sum_{i = 1}^{K} \phi_i \mathcal{N}(\mu_i, \Sigma_i).
    \label{eq:app_gmm}
\end{equation}
Making the standard assumption that dihedral angles are uncorrelated from bond lengths and bond angles, for a residue with $z$-matrix  $Z \in \RR^{k\times 4},$ we train a Gaussian Mixture Model (GMM) to generate bond lengths and bond angles and a second Gaussian Mixture Model (GMM) to generate dihedral angles. During reconstruction, we concurrently draw samples from the coupled bond length and bond angle generator $\rho^{\textrm{GMM}_{l,a}}$ and the dihedral generator $\rho^{\textrm{GMM}_{d}}.$  Given that the joint distribution of bond lengths and bond angles is unimodal, for $\rho^{\textrm{GMM}_{l,a}},$ the number of components $K = 1,$ and $\mu_i \in \RR^{2k}$ and $\Sigma_i \in \RR^{2k\times 2k}.$ For $\rho^{\textrm{GMM}_{d}},$ the number of components is defined on a per-residue basis and was manually identified (Table~\ref{tab:n_comp}).

The GMMs are trained using a standard Expectation-Maximization (EM) algorithm using the $\texttt{scikit-learn}$ package \cite{scikit-learn}. This optimization process is computationally efficient, with the largest residues taking less than ten CPU minutes. Importantly, we can interpret each Gaussian component of $\rho^{\textrm{GMM}_{d}}$ as corresponding to a particular rotameric state, with $\mu_i$ defined as the mean dihedral coordinate of that state, $\Sigma_i,$ the covariance matrix of that state and $\phi_i,$ the relative proportion of that rotamer state relative to other rotamer states. We include the distributions of select dihedral angles and their components in Figure~\ref{fig:app_dihedral_gmm}.

\section{Computational details for transformer}
\label{app:trans}
We use the implementation of the transformer model \cite{vaswani_attention_2017} available in \texttt{PyTorch}. Our approach for the transformer architecture and training is similar to general implementations, with a few small differences, which we note below. The transformer relies on an encoder-decoder framework, where an input backbone sequence is encoded to autoregressively decode an output rotamer sequence. We consider the sequence of backbone dihedrals and residue identities as our input sequence and the sequence of components of the GMM (see Section~\ref{app:gmm}) to be the output sequence. See Table~\ref{tab:trans_hyp} for relevant hyperparameters.

\subsection{Backbone encoder}

\begin{table*}

\centering
\begin{tabular}{|p{7cm}||p{3cm}|p{3cm}|}
 \hline
 \multicolumn{3}{|c|}{Hyperparameters} \\
 \hline
 Hyperparameter &Chignolin&AR-IDR\\
 \hline
 Number of Encoder Layers & 6  & 6\\
 Number of Decoder Layers & 6  & 6\\
 Number of Attention Heads & 8  & 8\\
 Number of Encoder Tokens (${n_{\rm tokens}^{\rm bb}}$) & 7  & 15\\
 Number of Decoder Tokens (${n_{\rm tokens}^{\rm rot}}$) & 67  & 67\\
 Embedding Dimension ($e_{\rm dim}$) & 256  & 256\\
 Optimizer & Adam  & Adam\\
 Learning Rate  & $1 \times 10^{-6}$    &$1 \times 10^{-6}$\\
 Number of Datapoints & 62M & 4M \\
 Batch Size & 512 & 512 \\
 \hline
\end{tabular}
\caption{Relevant hyperparameters for transformer training}
\label{tab:trans_hyp}
\end{table*}

\label{app:back_enc}
We represent the backbone using a sequence of categorical and continuous features corresponding to each residue. Each residue has a single categorical feature corresponding to its amino acid identity, which is passed into an \texttt{Embedding} layer. Because the two systems we investigate here do not individually contain all 20 amino acids, in our \texttt{Embedding} lookup table, we only include amino acids that are present in the protein.
The $\phi$ and $\psi$ dihedral angles contain rich information about the backbone and furthermore are strongly correlated with $\chi_1,$ the dihedral angle associated with the $\gamma-$Carbon (for appropriate residues). With that in mind, we consider a continuous tokenization that employs the $\phi$ and $\psi$  angles. To avoid issues with the periodicity of dihedral angles (e.g. $\pi$ and -$\pi$ are equivalent), we use the following continuous tokenization:
\begin{equation}
\begin{bmatrix}
\sin(\phi_1) & \sin(\phi_2) & \cdots & \sin(\phi_m) \\
\cos(\phi_1) & \cos(\phi_2) & \cdots & \cos(\phi_m) \\
\sin(\psi_1) & \sin(\psi_2) & \cdots & \sin(\psi_m) \\
\cos(\psi_1) & \cos(\psi_2) & \cdots & \cos(\psi_m)
\end{bmatrix},
\end{equation}
where $m$ denotes the protein sequence length.
Given a sequence of length $m$ and an embedding dimension of size $e_{\rm dim},$ the sequence of categorical features is embedded---via the $\texttt{Embedding}$ lookup table---according to a transformation $\NN^{m} \to \RR^{m\times e_{\textrm{dim}}/2},$ while the sequence of continuous features is embedded according to a standard \texttt{Linear} layer $\RR^{m\times 4} \to \RR^{m\times e_{\textrm{dim}}/2}$.  Finally, the categorical and continuous embeddings are concatenated to produce an embedding of size $\RR^{m\times e_{\textrm{dim}}}$, which is then passed through a standard sinusoidal positional encoder \cite{vaswani_attention_2017} and into the encoder block of the transformer.  We refer the reader to our code for additional details.

\subsection{Rotamer decoder}
\label{app:rot_dec}
Using the input sequence of backbone features, we aim to decode a rotamer sequence consisting of Gaussian component indices that can subsequently be used to generate rotamer states.  This rotamer sequence can be defined as  $(C_1, C_2 \dots C_m)$  where $C_j \in [1, K_{\rm max}]$ denotes the index of the Gaussian component for $\rho_j^{\rm GMM}$ (i.e.\ the dihedral GMM of the $j$th residue) and $K_{\rm max} = \max(K_1, K_2, \dots K_n)$ corresponds to the max number of total components across all dihedral GMMs. This last point is an important one; for residue $j$, we do not constrain the component index to be less than $K_j$ but instead constrain it to be less than $K_{\rm max}.$ While this results in a weaker inductive bias, it allows for a much simpler architecture and training process. Finally, we prepend a start-of-sequence token and append an end-of-sequence token to the sequence $(\rm{SOS}, C_1, C_2 \dots C_m, \rm{EOS}).$ 

Denoting the entire input backbone sequence $\xb$---where $\xb$ consists of both categorical and continuous features---and the rotamer sequence as $\yb,$  we can train the transformer according to the following cross-entropy loss
\begin{equation}
    \mathcal{L}({\theta}) = -\sum_{j=1}^{m+1} \textrm{CE}\bigl(\yb_{j + 1}, f(\xb, \yb_{1:{j}};\theta)\bigr),
\label{eq:app_ce}
\end{equation}
where CE is the cross-entropy loss, $f$ is the transformer, and $\theta$  is the parameters of the transformer neural network. In practice, the computation of this loss can be parallelized across the sequence by applying an appropriate masking operation. We refer the reader to our code for further implementation details.

\begin{algorithm}[H]
\caption{Generating a sequence of components}
\label{alg:gen_comp}
\begin{algorithmic}[1]
    \State{Initialize transformer $f$ and load trained parameters $\theta$}
    \State{Prepare backbone encoder sequence $\xb$}
    \State{Defined $y_{\rm next} = \rm SOS$}
    \State{Define $\yb = []$}
    \State{Append $y_{\rm next}$ to $\yb$}
    
    \While {$y_{\rm next} \neq \rm EOS$}
        \State{$y_{\rm pred} = f(\xb, \yb; \theta)$}
        \State{Sample $y_{\rm next}$ from $y_{\rm pred}$}
        \State{Append $y_{\rm next}$ to $\yb$}
    \EndWhile
\end{algorithmic}
\end{algorithm}

\subsection{Predicting a sequence of components}
During inference time, conditioned on a backbone structure, we seek to generate a sequence of rotamer components that can be subsequently used to sample rotamers across the full protein. Denoting the sequence of backbone features $\xb$ (see Appendix~\ref{app:back_enc},~\ref{app:rot_dec} for further details), our initial rotamer sequence $\yb_{1:1} =  (\rm SOS),$ where SOS represents the start-of-sequence token. As in Appendix~\ref{app:rot_dec}, the backbone sequence is passed into the encoder block of the transformer and the rotamer sequence is passed into the decoder block of the transformer. The output of the transformer $f(x, y_{1:{j}};\theta) \in \RR^{n_{\rm tokens}^{\rm rot}}$ is a multinomial distribution, where $P(C_j = k) = f(\xb, \yb_{1:{j}};\theta)_k.$ While many strategies exist for autoregressively generating a sequence, here, we select a straightforward strategy; for every element in the sequence, we randomly sample $C_j$ according to the multinomial distribution generated by the transformer $f(\xb, \yb_{1:{j}};\theta) \in \RR^{n_{\rm tokens}^{\rm rot}}.$ This process occurs until an end-of-sequence token is sampled, at which point the sequence terminates. See Algorithm~\ref{alg:gen_comp} for a summary of the sequence generation process.

The number of components is residue-specific; smaller residues generally have a smaller number of components, while larger residues (e.g.\ those with longer alkyl chains) contain a larger number of components. The number of tokens for the decoder ${n_{\rm tokens}^{\rm rot}}$ is fixed and corresponds to the maximum number of components across all dihedral GMMs. During inference, if, for any single sequence $j,$ a sampled $C_j > K_j,$ where $K_j$ is the number of components for the dihedral GMM of the $j$th residue, the sequence is considered invalid. For Glycine, we do not need to backmap any atoms but still include it in our sequence and so assign it a separate token. As a result, including the special Glycine token, the SOS token, and the EOS token, $n_{\rm tokens}^{\rm rot} = C_{\rm max} + 3$

\begin{figure}
    \centering
    \includegraphics[width=0.8\linewidth]{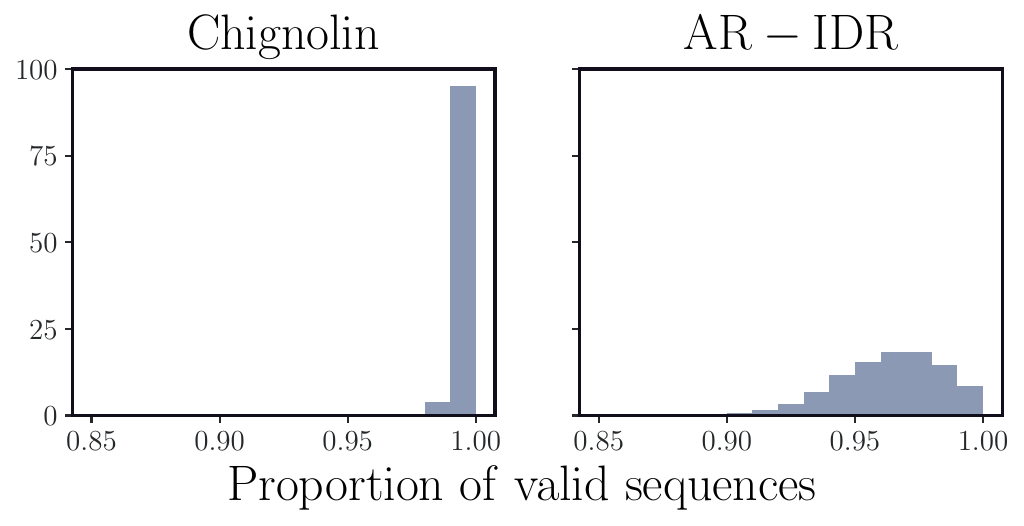}
    \caption{Proportion of valid sequences across all backbones in the test sets of Chignolin and AR-IDR. For a given backbone, the median sequence validity is 1.0 for Chignolin and 0.97 for AR-IDR.}
    \label{fig:frac_val_seq}
\end{figure}

Furthermore, in many tasks transformers are deployed to generate sequences of variable lengths. In the task we consider here, we are interested in generating sequences of a fixed length as the number of residues for the proteins we consider is fixed. As noted, we grow a sequence autoregressively until the end-of-sequence (EOS) token is reached. If the length of the generated sequence (including the SOS and EOS token) is not $m + 2,$ the sequence is considered invalid. We plot the proportion of valid generated sequences across all backbones in the test in Figure~\ref{fig:frac_val_seq}. The sequence validity is high across all backbones demonstrating our model can capture the underlying structure of the rotamer sequence.

\section{Details for Chignolin and AR-IDR}
To train the transformer according to Eq.~\ref{eq:app_ce}, we of course need access to a dataset. The setting we consider here is one in which we have access to a collection of backbone configurations, which can be obtained by coarse-grained simulations. Here, we select structures from all-atom MD simulations and coarse-grain these structures to a backbone resolution. We backmap each of these backbone structures by sampling from the residue-specific GMMs $\rho^{\textrm{GMM}_{d}}_j$ and  $\rho^{\textrm{GMM}_{l,a}}_j,$ in an uncoupled fashion. This strategy enables us to rapidly create a large number of independent structures. Of course, this uncoupled sampling strategy will result in many high-energy unphysical configurations. Furthermore, many structures will have a high energy, but a structure that is quantitatively very similar to a low energy structure. Often, a small change in the conformation (~0.01 Å) can dramatically reduce the energy of a structure by many $k_{\rm B}T$. With this in mind, we perform a short relaxation step of all structures. Finally, we select all configurations below -50 $k_{\rm B}T$ and use these configurations for our dataset.

We note again that, while the configurations we select are comparatively low-energy and have statistical validity under the true Boltzmann distribution, our dataset is \textit{not} Boltzmann-distributed. Furthermore, we note that technically we can---under a relaxation scheme that uses a reversible integrator---compute a reweighting factor exactly and ensure our samples are Boltzmann-distributed. However, the entropy of our proposal scheme via uncoupled GMMs is much greater than that of the true rotameric density, and this results in reweighting factors dominated by a handful of terms. Practically under such a reweighting scheme, most configurations would get down-weighted, greatly diminishing sample diversity in a reweighted dataset. Because of this limitation, we relax the requirement that our training set be Boltzmann distributed, with the goal of carrying out a reweighting step at a downstream step (see Section~\ref{sec:outlook}).

\label{app:prot}
\subsection{Chignolin}
\label{app:prot_chig}
For Chignolin, we selected 50000 backbone configurations from a long MD simulation \cite{husic_coarse_2020}. For each of these backbone configurations, we generated 10000 possible states via uncoupled rotameric sampling, carried out a relaxation step, and selected all configurations below -50 $k_{\rm B}T,$ where T = 300K. The relaxation step consisted of 5 steps of Verlet integration with dt=0.001 ps at 300K using an implicit solvation model. We used a train/validation/test, where the split was done at the backbone level so that no backbone configuration was present in two or more sets. This led to a training dataset of 62 million configurations, where for each backmapped configuration, we recorded the relevant backbone information and the components of the dihedral GMMs that were used to generate the configuration.

\begin{figure}
    \centering
    \includegraphics[width=1.0\linewidth]{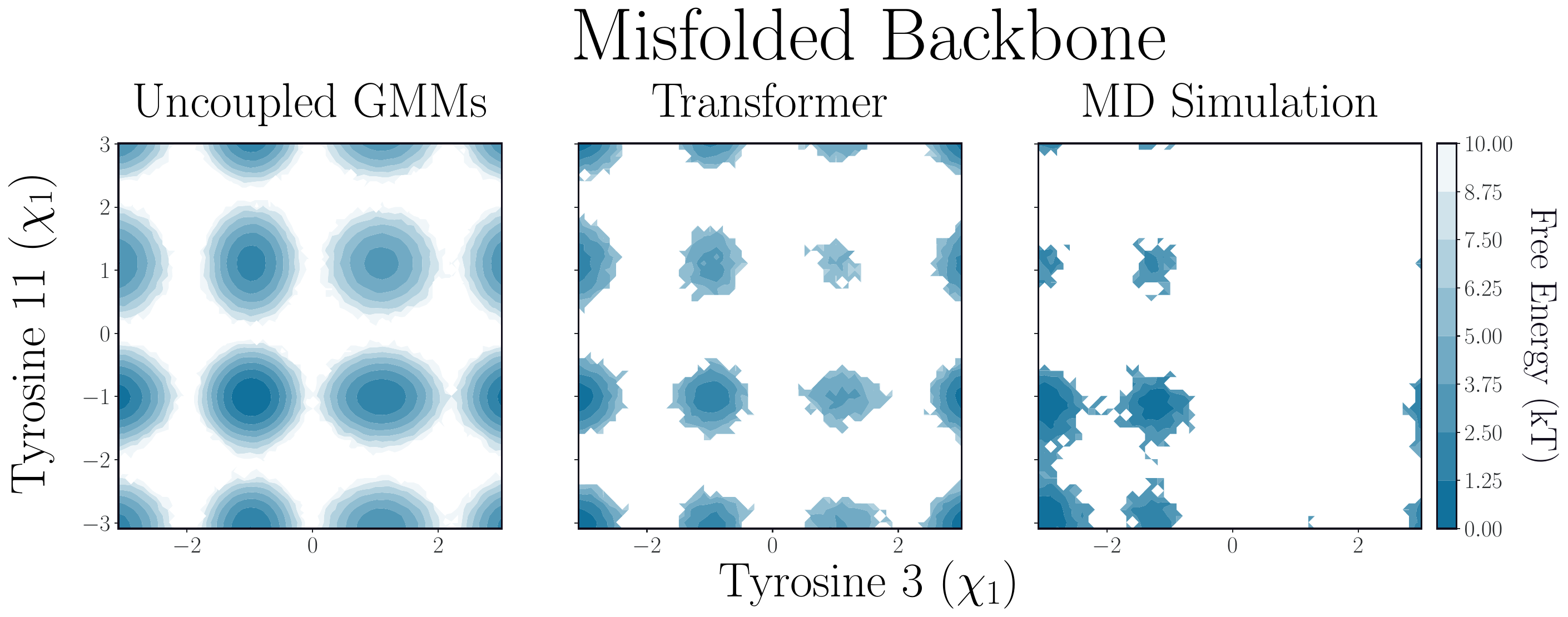}
    \includegraphics[width=1.0\linewidth]{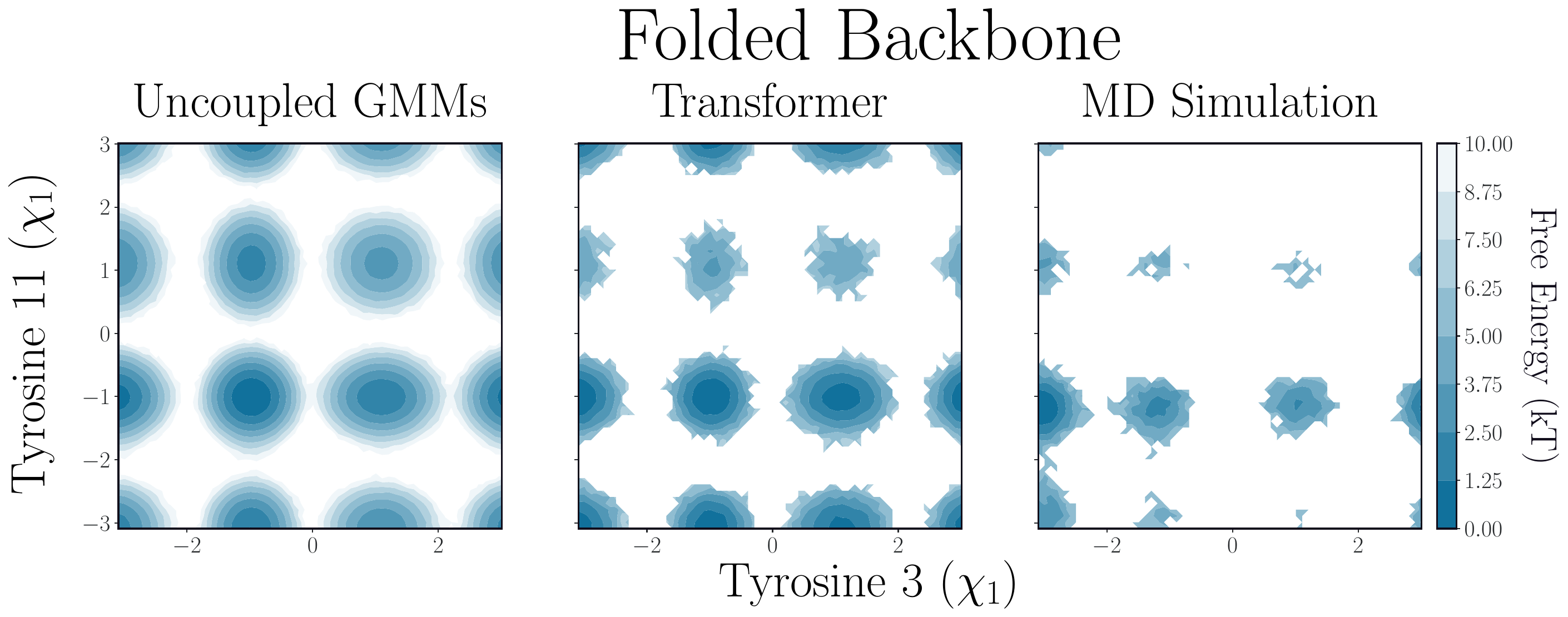}
    \includegraphics[width=1.0\linewidth]{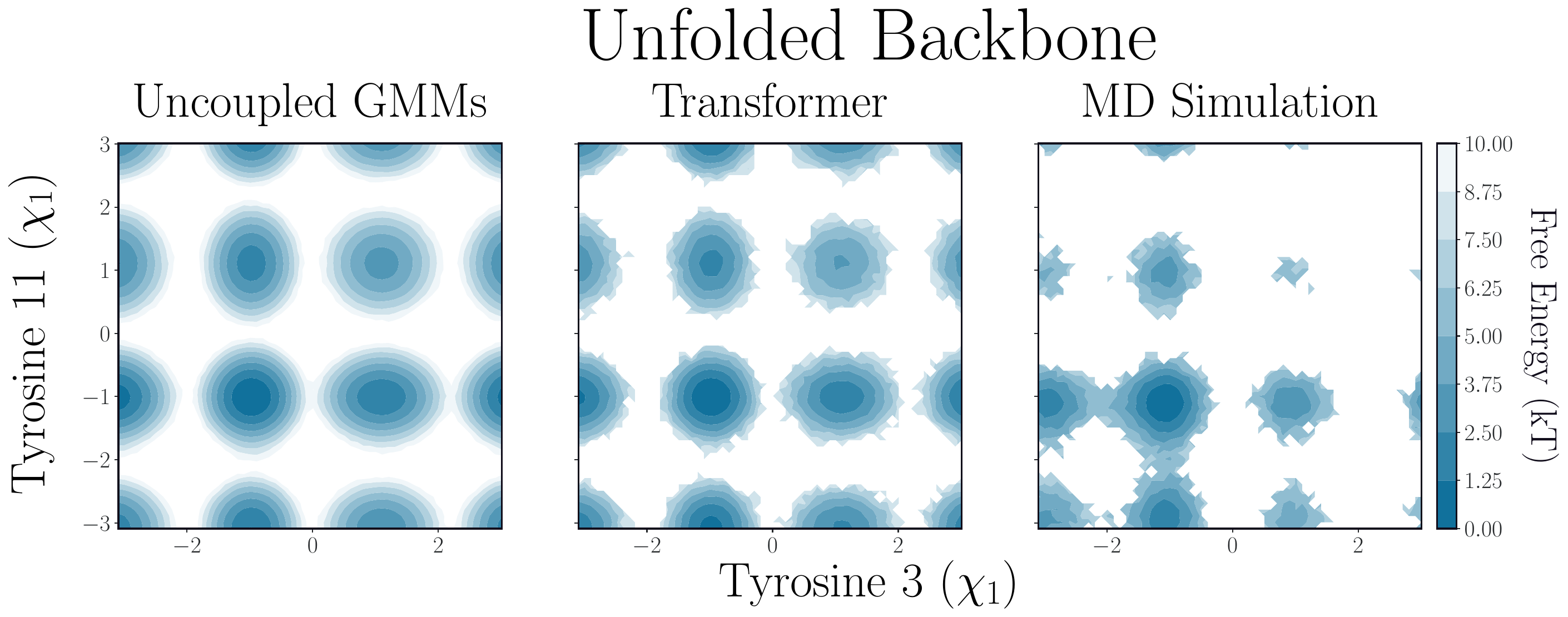}
    \caption{For transformer-coupled sampling and MD simulation, as the hairpin interior becomes more obstructed (unfolded $\to$ folded $\to$ misfolded), the rotamer entropy decreases, and more states become disallowed. Sampling uncoupled generators does not recapitulate this trend.}
    \label{fig:all_tyr_tyr_dihedrals}
\end{figure}

To be consistent with the indexing in the trajectory obtained from \cite{husic_coarse_2020}, the ACE cap is indexed at 1, so TYR2 corresponds to the first non-cap residue, TYR3 the second non-cap residue, and TYR11 the tenth and terminal non-cap residue. We plot the distribution of all rotamer states across all backbones in Figure~\ref{fig:all_tyr_tyr_dihedrals}.

\subsection{AR-IDR}
\label{app:prot_ar}
For the intrinsically disordered region of the androgen receptor (AR-IDR), we selected 57144 backbone configurations from a long MD simulation \cite{zhu_small_2022a}. For each of these backbone configurations, we generated 50000 possible states via uncoupled rotameric sampling, carried out a relaxation step, and selected all configurations below -50 $k_{\rm B}T,$ where T = 300K. The relaxation step consisted of 20 steps of an overdamped Langevin integration with dt=0.00002 ps and a friction coefficient of $\gamma = 100.0 $ $\rm{ps}^{-1}$ at 300 K using an implicit solvation model. We used a train/validation/test, where the split was done at the backbone level so that no backbone configuration was present in two or more sets. This led to a training dataset of 4 million configurations, where for each backmapped configuration, we recorded the relevant backbone information and the components of the dihedral GMMs that were used to generate the configuration.

For the AR-IDR, the original trajectory considered a 56 residue region (L391-G446 region) of the N-terminal transactivation domain; for simplicity, we reindex that in our analysis such that the ACE cap is indexed at 0. So, Trp7, Trp43, and Phe47 in our analysis corresponds to Trp397, Trp433, and Phe437 in \cite{zhu_small_2022a}. 

To define helicity, we used the order parameter $S_{\alpha}$ in \cite{zhu_small_2022a}, which measures discrepancy from an ideal helix using a sliding window of length six residues. For Phe7, we consider the $S_{\alpha}$ computed from the four windows starting at residues 3-7. For Phe7, we categorize the backbone as weakly helix ($S_{\alpha} < 0.5$) and strongly helix ($S_{\alpha} \geq 0.5)$. For Trp43 and Phe47, which reside in the same helix-forming region, we measure the helicity from the five windows starting at residues 38-43. With Phe7, we defined four classes for helicity: no helix ($S_{\alpha} < 0.15$), weakly helix ($0.15 \leq S_{\alpha} < 0.4$), moderately helix ($0.4 \leq S_{\alpha} < 0.7$), and strongly helix ($S_{\alpha} \geq 0.7$).

\begin{figure}
    \centering
    \includegraphics[width=\linewidth]{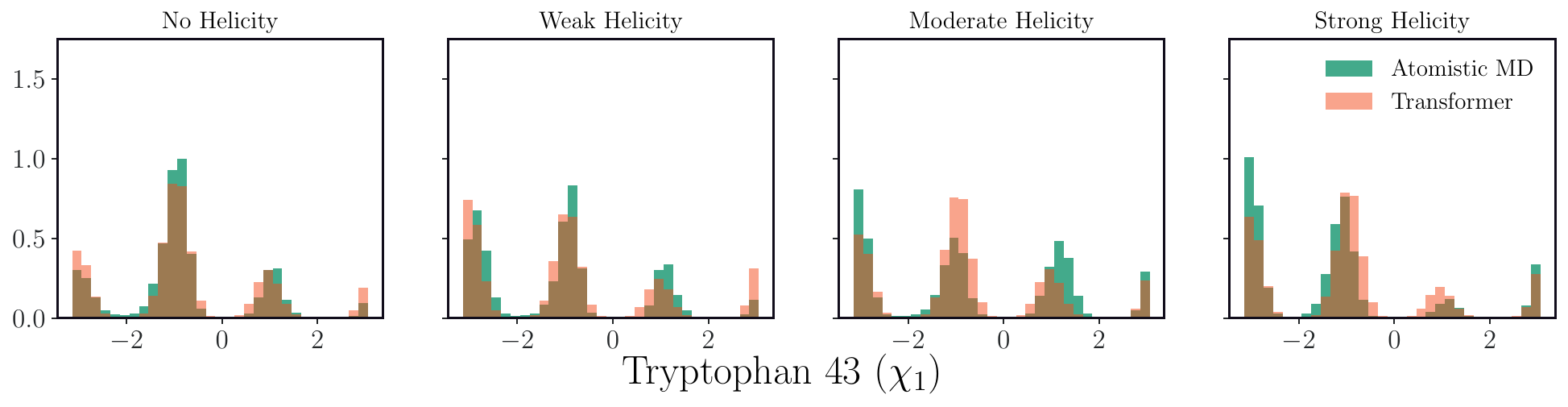}
    \includegraphics[width=0.5\linewidth]{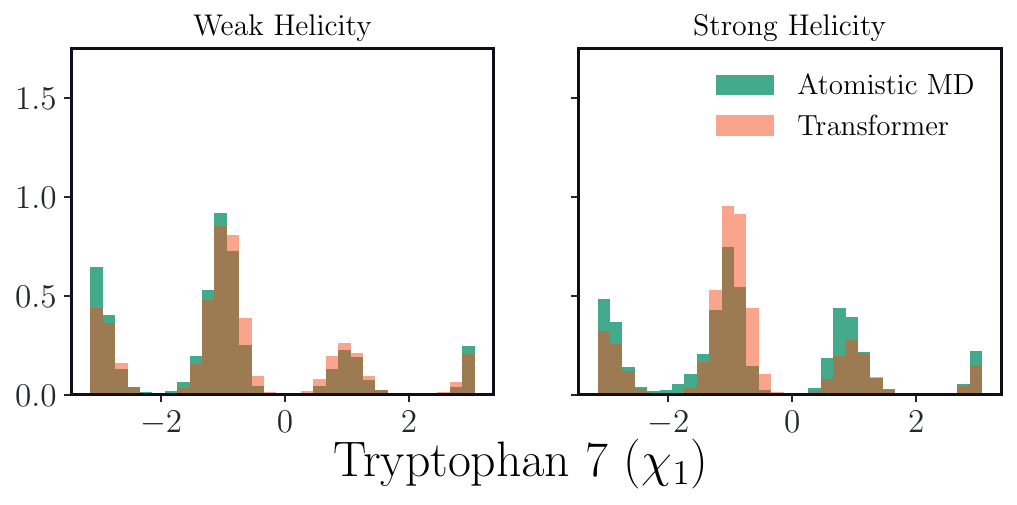} 
    \caption{Effect of backbone helicity on rotamer state of Trp43 and Trp7. Transformer-coupled sampling is in strong agreement with rotamer states observed during MD simulation.}
    \label{fig:app_helicity}
\end{figure}

We analyze the effect of helicity on Trp7 and Trp43 in Figure~\ref{fig:app_helicity}. For Trp43 as the degree of helicity increases, the rotamer state centered at $\pi$ becomes more probable under both the transformer-coupled scheme and the long MD simulation. For Trp7, we observe that the rotamer density is weakly dependent on the degree of helicity in MD simulations and this is a trend that is also recapitulated in transformer-informed sampling.

\section{Comparison to other backmapping methods}
\label{app:benchmark}
There has been recent interest in designing data-driven methods for protein backmapping from coarse-grained models \cite{yang_chemically_2023, jones_diamondback_2023, liu_backdiff_2023}. To judge how the transformer-coupled backmapping approach introduced here compares to select existing state-of-the-art data-driven backmapping strategies, we carry out backmapping for two proteins deposited in the Protein Ensemble Database (PED)~\cite{ghafouri_ped_2024} and evaluate conventionally used metrics on the backmapping. Specifically, we compare our method against GenZProt\cite{yang_chemically_2023}, BackDiff~\cite{liu_backdiff_2023}, and Torsional Diffusion (TD)~\cite{jing_torsional_2023} and summarize the results in Table~\ref{tab:backmapping_comparison}. As in Ref.~\cite{liu_backdiff_2023}, we carry out backmapping for the intrinsically disordered proteins PED00011 \cite{sanchez-martinez_application_2014} and PED00151 \cite{fuertes_decoupling_2017}. 

In an effort to ensure a fair comparison with other backmapping strategies, we assume \textit{a priori} that we have access to only the dataset present in the PED---5926 frames for PED00011 and 9746 frames for PED00151---and the corresponding simulations of all tripeptides relevant for each of the proteins. We note that this is a departure from the setting that this method is designed for and of the analysis for Chignolin and AR-IDR. In that analysis, we did not have access to all-atom configurations and instead generated all-atom data using rotamer generators; however, for PED00011 and PED00151, we do have access to all-atom configurations and do not generate further all-atom data.

For each configuration, we identified the relevant rotamer state by identifying the GMM component that had a maximum likelihood of a specific side chain configuration. Here, each GMM was trained on tripeptide simulations run using an Amber force field, which we note is different from the force field used for the simulations for PED00011 and PED00151. We carried out an 80/10/10/ train/validation/test-set split and carried out all analyses on the test set. For each backbone configuration in the test set, we backmap 100 all-atom configurations.

We evaluate the backmapping using 4 metrics: the minimum Root Mean Squared Distance to a reference structure $\text{RMSD}_\text{min}$, the steric clash ratio (SCR), the Mean Squared Error of the center of mass of generated side chains from a reference structure SCMSE$_{\text{min}}$, and the generative diversity score (DIV). A lower DIV score correlates with a larger diversity of generated structures and a lower SCR corresponds to the successful generation of physically plausible structures. While RMSD is a useful metric for many contexts, it is less useful as a metric for evaluating a backmapping strategy. The reference structure is merely a static representation of a highly dynamic structure with numerous side chain conformations. Observing a low RMSD (and the closely related SCMSE) only demonstrates that a backmapping approach closely recapitulates a structure in the dataset but is ultimately uninformative on how well a backmapping technique samples the underlying rotamer density. Still, we include these metrics for completeness, and we refer the reader to Ref.~\cite{liu_backdiff_2023} for further detail on the computation of each of these metrics.


\begin{table}[h!]
\begin{center}
\begin{tabular}{cccc}
\toprule & Method & PED00011 & PED00151  \\
\midrule  & Transformer & $1.831(0.081)$ & $1.932 (0.083)$\\
& BackDiff (fixed) & $0.415(0.107)$ &$0.526 (0.125)$ \\
$\text{RMSD}_\text{min}$ ($\text{\AA}$) & GenZProt & $1.392(0.276)$ &$1.246(0.257)$ \\
& TD & $1.035(0.158)$ &$1.253(0.332)$ \\
\midrule & Transformer & $0.151(0.033)$ &$0.042(0.019)$\\
& BackDiff (fixed) & $0.100(0.035)$&$0.105(0.063)$\\
SCR ($\%$)& GenZProt & $0.408(0.392)$
&$0.647(0.384)$\\
& TD & $0.356(0.303)$ &$0.452(0.187)$\\
\midrule & Transformer & $0.189(0.029)$ &$0.217(0.029)$ \\
& BackDiff (fixed) & $0.045(0.008)$&$0.049(0.021)$ \\
SCMSE$_{\text{min}}$ ($\text{\AA}^2$)& GenZProt & $1.225(0.121)$ &$1.340(0.182)$ \\
& TD & $1.134(0.125)$ & $1.271(0.158)$\\
\midrule & Transformer & $0.046(0.029)$ &$0.077(0.031)$ \\
& BackDiff (fixed) & $0.045(0.027)$ &$0.072(0.034)$ \\
$\text{DIV}$& GenZProt & $0.453(0.241)$ &$0.527(0.185)$ \\
& TD & $0.128(0.064)$ & $0.146(0.049)$\\
\bottomrule
\end{tabular}
\end{center}
\caption{Transformer-coupled sampling produces results with a larger diversity score and lower steric clash ratio. Results for compared methods are reproduced from \cite{liu_backdiff_2023}.}
\label{tab:backmapping_comparison}
\end{table}

While the $\text{RMSD}_\text{min}$ we compute are comparatively higher than the other 3 methods, we are able to generate structures for which the center of mass of each side chain is close to that of the reference structure as evidenced by the comparatively low SCMSE$_{\text{min}}.$ More importantly, we succeed in generating a high diversity of structures with limited steric clashes; our quantitative evaluation is competitive with or better than the best-performing method.

In the dataset for PED00011 and PED00151, for many of the side chains, the distribution of $\chi_1$ dihedral angles has significant probability mass outside the three canonically observed rotamer states (e.g.\ Fig.~\ref{fig:chigtyr}~a), which is likely a product of both the unique force fields used and the reweighting of configurations with respect to experimental measurements. As such, the rotamer generators that we use are not perfectly tailored to the PED dataset. Still, we can carry out backmapping and we observe results that are competitive with or better than existing state-of-the-art backmapping methods. More importantly, we stress that the method introduced here is designed for settings in which all-atom data is limited or not present. While the methods we compare to can be adapted to protein systems with limited all-atom data via a transfer learning strategy, it is unclear how the diversity in sampling approaches (e.g.\ different force fields) and limited number of protein systems (92) in the PED will impact a transfer learning strategy \cite{liu_backdiff_2023}.

\end{document}